\documentclass[journal]{IEEEtran}
\usepackage{cite}
\usepackage{amsmath,amssymb,amsfonts}
\usepackage{algorithmic}
\usepackage{graphicx}
\usepackage{textcomp}
\usepackage{xcolor}
\usepackage{makecell}
\usepackage{algorithm}
\usepackage{algorithmic}
\usepackage{upgreek}
\usepackage{stfloats}

\def\BibTeX{{\rm B\kern-.05em{\sc i\kern-.025em b}\kern-.08em
    T\kern-.1667em\lower.7ex\hbox{E}\kern-.125emX}}

\begin{document}

\bibliographystyle{IEEEtran}

\title{Mutual Interference Mitigation \\ for MIMO-FMCW Automotive Radar}
\author{Sian Jin, Pu Wang, Petros Boufounos, Philip V. Orlik, Ryuhei Takahashi, and Sumit Roy
\thanks{S. Jin is a Post-doctoral Research Associate, Princeton University, Princeton, NJ, USA. 
P. Wang, P. Boufounos and P. Orlik are with Mitsubishi Electric Research Laboratories (MERL), Cambridge, MA, USA. 
R. Takahashi is with Mitsubishi Electric Information Technology R\&D Center, Ofuna, Kamakura City, Japan.
S. Roy is with University of Washington, Seattle, WA, USA. }
\thanks{The work of S. Jin was partially conducted during his internship at MERL and his Ph.D. studies at University of Washington.}
\thanks{Part of this paper will be presented in 2023 ICASSP~\cite{ICASSP23}.}
}

\newtheorem{Thm}{Theorem}
\newtheorem{Lem}{Lemma}
\newtheorem{Cor}{Corollary}
\newtheorem{Def}{Definition}
\newtheorem{Exam}{Example}
\newtheorem{Alg}{Algorithm}
\newtheorem{Sch}{Scheme}
\newtheorem{Prob}{Problem}
\newtheorem{Rem}{Remark}
\newtheorem{Proof}{Proof}
\newtheorem{Asump}{Assumption}
\newtheorem{Subp}{Subproblem}
\newtheorem{prop}{Proposition}

\maketitle
\date{}

\begin{abstract}
This paper considers mutual interference mitigation among automotive radars using frequency-modulated continuous wave (FMCW) signal and multiple-input multiple-output (MIMO) virtual arrays. For the first time, we derive a general interference signal model that fully accounts for not only the time-frequency incoherence, e.g., different FMCW configuration parameters and time offsets,  but also the slow-time code MIMO incoherence and array configuration differences between the victim and interfering radars. Along with a standard MIMO-FMCW object signal model, we turn the interference mitigation into a spatial-domain object detection under incoherent MIMO-FMCW interference described by the explicit interference signal model, and propose a constant false alarm rate (CFAR) detector. More specifically, the proposed detector exploits the structural property of the derived interference model at both \emph{transmit} and \emph{receive} steering vector space. We also derive analytical closed-form expressions for probabilities of detection and false alarm. Performance evaluation using both synthetic-level and phased array system-level simulation confirms the effectiveness of our proposed detector over selected baseline methods.
\end{abstract}

\section{Introduction}
Advanced driver assistance systems (ADAS) and autonomous driving require a high-resolution environment perception system capable of detecting and identifying stationary (e.g., buildings, trees, and guardrails) and dynamic (e.g., vehicles and pedestrians) objects reliably in all weather conditions. 
Compared with other perception sensors such as cameras and LiDAR, radar offers the potential for operating in adverse weather and night-time conditions at lower cost and processing overhead~\cite{IntMag19}.

Current automotive radars widely adopt frequency-modulated continuous wave (FMCW) techniques~\cite{SAM22,ICASSP23,IntMag19,Torlak, BilikLongman19, HakobyanYang19, WangMillar18, Wang19, WangBoufounos20}, since it enables receivers with low sampling rates while harnessing large sweep frequency bands for high resolution in range. On the other hand, they are limited in use for high-resolution perception tasks due to poor angular resolution, particularly in the elevation domain. To increase the angular resolution, automotive radar chip vendors take various approaches to form a large aperture for highly directional beams. Mechanically scanned FMCW radars, e.g., Navtech CTS350-X,  have been used to collect $360^{\circ}$ bird's-eye view (BEV) radar images in the range-azimuth domain but without the Doppler velocity~\cite{radiate}.  Synthetic aperture radar (SAR) techniques create high-resolution two-dimensional images of the scene by coherently combining returned radar waveforms with the assumption of known ego vehicle motion~\cite{MostajabiWang20}. Multiple-input multiple-output (MIMO) radar 
is another cost-efficient approach to form a large virtual array with a reduced number of transmitting (Tx) and receiving (Rx) antennas and radio frequency (RF) chains. To achieve this, one needs to separate the corresponding waveform to each transmitter at each receiver, provided that the transmitting waveforms from different Tx antennas can be separable or orthogonal. Orthogonal MIMO signaling schemes can be realized in time-division multiplexing (TDM), frequency-division multiplexing (FDM), and Doppler-division multiplexing (DDM) (also referred to as slow-time MIMO) modes~\cite{Rao17, WangBoufounos20, SunPetropulu20}. As of today, the combined MIMO-FMCW automotive radar has been commercialized by chip vendors to achieve hundreds and even thousands of virtual channels in the azimuth and elevation domains \cite{Rao17b, OchPfeffer18}. 

\begin{figure}[t]
\begin{center}
  {\resizebox{9cm}{!}{\includegraphics{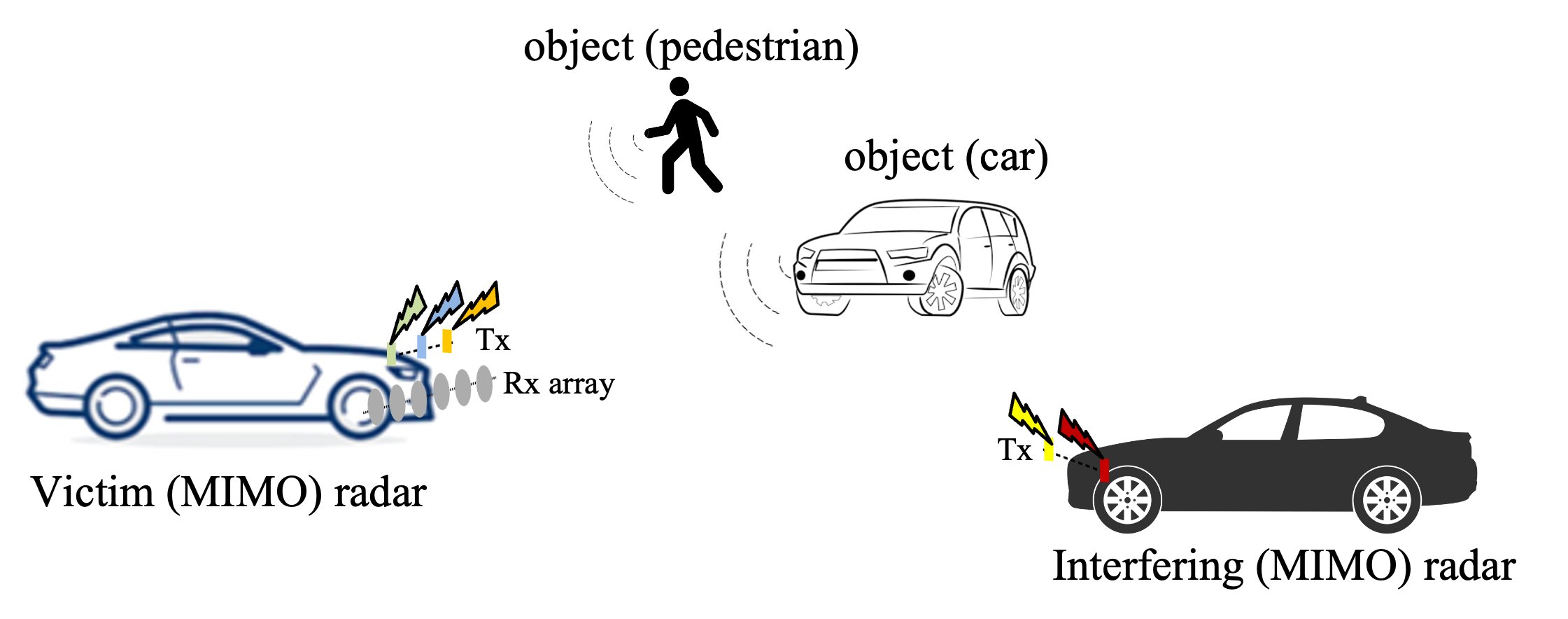}}}
         \caption{Illustration for mutual interference mitigation for MIMO-FMCW automotive radar, where both victim and interfering vehicles use MIMO transmitter and receiver arrays to transmit and receive waveform.}\label{fig:illustration}
\end{center}
\end{figure}

When multiple automotive radars operate in the same regulated frequency bands, e.g., $76-81$ GHz, it is anticipated that mutual radar interference becomes a serious issue, as shown in Fig.~\ref{fig:illustration}. Mutual interference mitigation has been considered for traditional FMCW radar and can be classified as: 
\begin{enumerate}
    \item Fast-time (range) domain:  interference-zeroing~\cite{zeroWindow,STFTCFAR,STFTZeroInterpolate}, sparse reconstruction~\cite{IMAT,Chirplets}, adaptive noise cancellers~\cite{ANC}, signal separation~\cite{separation}, fast-time-frequency mode retrieval~\cite{SianFSST}, and fast-time neural networks~\cite{STFTDL,STFTDL2};
    \item Slow-time (Doppler) domain: waveform randomization~\cite{SianJSTSP,Randomisation}, ramp filtering~\cite{rampFiltering}, and slow-time neural network~\cite{slowTimeGAN}; 
    \item Joint range-Doppler domain:  neural network based denoisers~\cite{RVAE,RVGAN,RVSCNN,RVQCNN};
    \item Communication-assisted scheduling, such as time-division multiple access~\cite{RadChat}, and chirp slope and frequency offset scheduling~\cite{HeathCoordination}.
\end{enumerate}

For MIMO-FMCW automotive radar, interference mitigation can be done in the MIMO code domain \cite{BoseTang21} but it requires additional communication and coordination between the victim and interfering radars. On the other hand, spatial-domain mitigation approaches were considered to make use of additional degrees of freedom in the antenna or beamspace domain. Initial efforts include receiver beamforming-based approaches~\cite{DBFAdaptive0,DBFAdaptive1,DBFAdaptive2,DBFAdaptive3,GIP}, null steering~\cite{DBF}, and linear constraints minimum variance (LCMV) beamforming~\cite{EuRADMIMO}. 

Different from all the above efforts, our approach is to understand the interference signal at the output of the range-Doppler and MIMO waveform separation of a standard automotive radar. With such a mathematical understanding of the interference signal, we are able to approximate the interference using a Kronecker subspace signal model in the spatial (object angles seen from both the transmitter and receiver) domain and turn the mutual interference mitigation into an object detection problem under a Kronecker subspace interference plus noise. Our contributions are summarized below:  
\begin{itemize}
    \item For the first time, we derive an \emph{explicit} signal model for the spatial-domain MIMO-FMCW interference  under the time-frequency incoherence,  the MIMO code incoherence, and the array configuration difference between the victim and interfering radars. 
    \item We also show that the derived interference signal model reduces to existing models used in the literature under special cases such as coherent interference, TDM-MIMO interference, and phased array interference. 
    \item We exploit the structure of \emph{both Tx and Rx} steering vectors of the incoherent interference. Particularly, we decompose the incoherent MIMO-FMCW interference into two orthogonal components: one is completely aligned with the object Tx steering vector, and the other is in its orthogonal complement subspace.
    \item We propose a generalized subspace-based (GS) detector that minimizes the variance of interference-plus-noise with known statistics after Rx beamforming, maintains a fixed gain at the object direction and cancels the residual incoherent interference. 
    \item  We derive closed-form analytical expressions of probabilities of false alarm and detection and confirm that the proposed detector has the property of constant false alarm rate (CFAR).
    \item  We further show analytical convergence to existing receive-subspace-based detectors (RS detector or null-steering detector) and the clairvoyant detector under certain conditions. 
    \item  We provide a comprehensive numerical comparison between the proposed detector and several baseline methods (including the state-of-the-art RS and LCMV detectors) using analytical performance curves, synthetic data, and more realistic data that accounts for element-wise array beampatterns.
\end{itemize}

Throughout this paper, we use $(\cdot)^T$ to represent transpose, use $(\cdot)^*$ to represent conjugate, and use $(\cdot)^H$ to represent conjugate transpose.
We use $\mathbf P_{\mathbf H} \triangleq \mathbf H (\mathbf H^H \mathbf H)^{-1} \mathbf H^H$ to denote the projection matrix projecting to the column space of $\mathbf H$.
We use $\mathbf P_{\mathbf H}^{\perp} \triangleq \mathbf I - \mathbf P_{\mathbf H}$ to denote the projection matrix projecting to the space orthogonal to the column space of $\mathbf H$.
$Q_1(x,y)$ denotes the Marcum Q-function of order $1$~\cite{Marcum}.
All indices are counted from $0$.

\begin{figure}[t]
\begin{center}
     \includegraphics[width=0.5\textwidth]{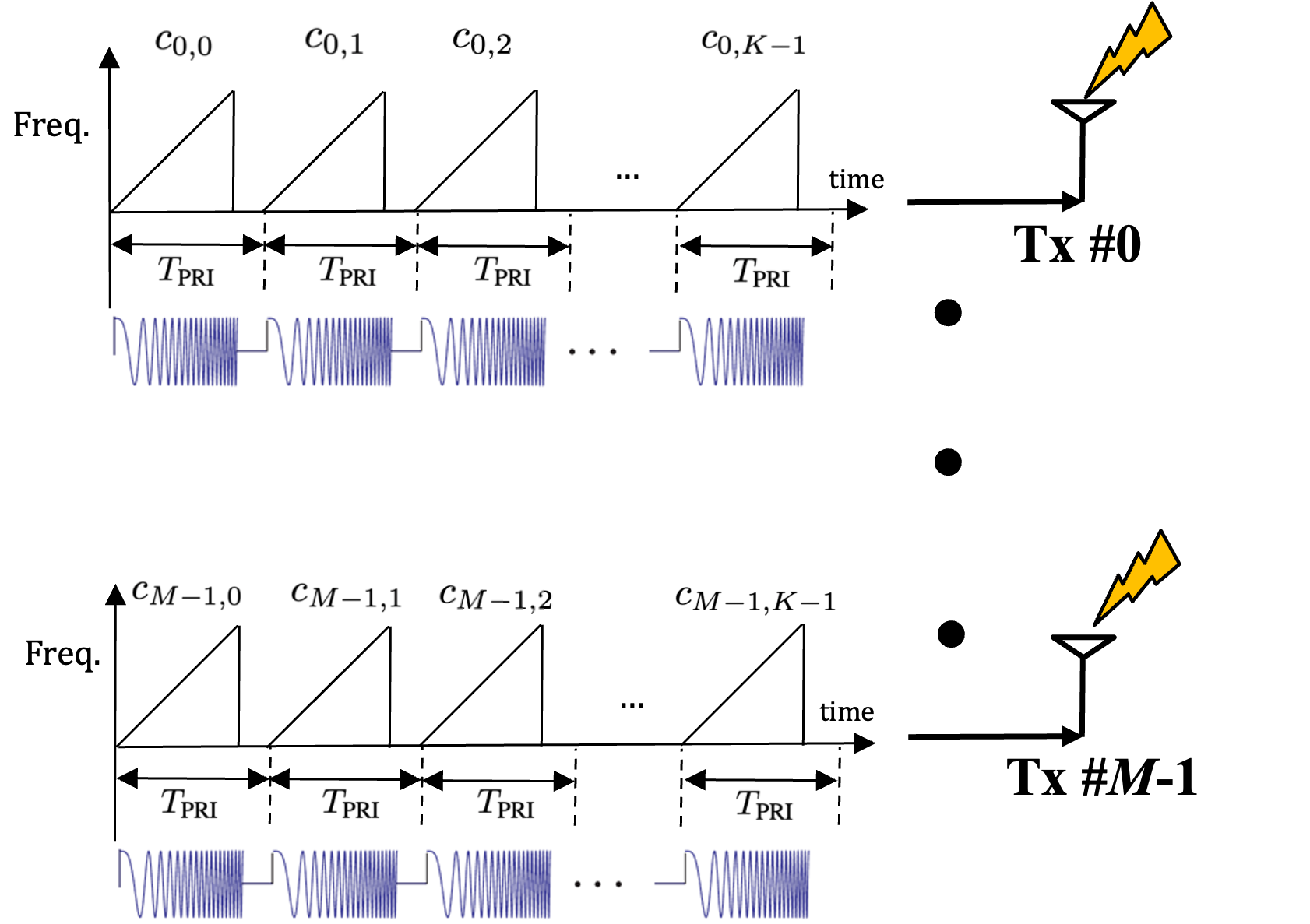}
\end{center}
    \caption{MIMO-FMCW waveforms with Tx-pulse code $c_{m,k}$ applied to the same source FMCW waveform. The Tx-pulse codes can vary depending on the operation mode: slow-time MIMO/DDM-MIMO (e.g., Hadamard or Chu sequences), TDM-MIMO (one-hot vectors), and phased array (all-one vectors with analog phase shifters).}
   \label{fig:fmcw}
   % \vspace{-0.1in}
\end{figure}
\section{Signal Model}\label{sec:model}

\begin{figure*}[t]
\begin{center}
\includegraphics[width=1.0\textwidth]{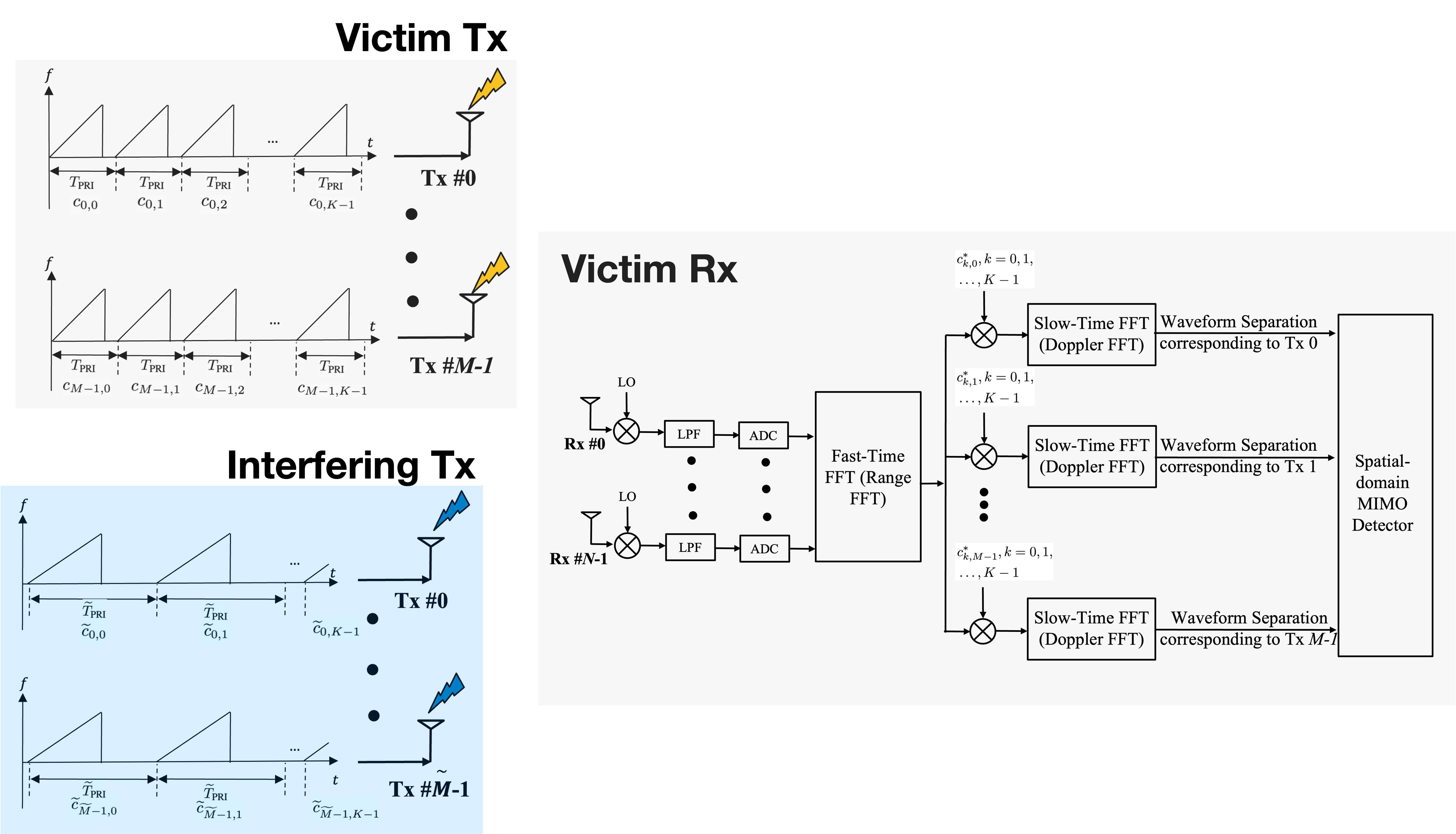}
\end{center}
    \caption{The receiver architecture (right) of a victim MIMO-FMCW automotive radar that captures both transmitted waveforms from its own transmitter (upper left) and an incoherent MIMO-FMCW interfering radar (lower left) with different FMCW configuration parameters, time offset, MIMO codes, and transmitter array configurations.}
   \label{fig:mimoBlockDiagram}
\end{figure*}

In the following, we briefly overview the object signal model, derive the interference signal model in more detail, and show the convergence of the derived interference model to existing results in various operation modes, e.g., coherent interference, phased array interference, and TDM-MIMO interference. 

\subsection{MIMO-FMCW Waveform} \label{sec: waveform}
As shown in Fig.~\ref{fig:fmcw}, we consider a victim radar of $M$ Tx antennas collocated with $N$ Rx antennas over $K$ pulses within a coherent processing interval (CPI).
The source FMCW waveform of the victim radar is 
\begin{align} \label{eqn: FMCW}
    s(t) = e^{j\pi \beta t^2} D_{0,T} (t),
\end{align}
where $\beta$ is the chirp rate, $T$ is the chirp duration, and
\begin{align}
D_{a,b} (t) = 
\begin{cases}
    1, & a \leq t \leq b \\
    0, & \text{otherwise}.
\end{cases}
\end{align}
The RF waveform on Tx antenna $m$ over $K$ pulses is~\cite{WangBoufounos20}
\begin{align} \label{eqn: chirpPulses}
    s_m(t) = \sum_{k=0}^{K-1} c_{k,m} s(t-k T_{\text{PRI}}) e^{j2\pi f_c (t -k T_{\text{PRI}})},
\end{align}
where $c_{k,m}$ is the Tx-pulse code on $m$-th Tx antenna and $k$-th pulse, $T_{\text{PRI}}$ is the pulse repetition interval of the victim radar and $f_c$ is the carrier frequency. In \eqref{eqn: chirpPulses}, the Tx-pulse codes may vary depending on the operation mode \cite{SunPetropulu20}:
\begin{itemize}
    \item slow-time MIMO/DDM-MIMO mode: the codes at Tx $m$ are chosen to achieve zero/low cross-correlation over transmitted antennas. One example is the use of binary Hadamard code where $c_{k,m}$ is taken from the columns of a Hadamard matrix of size $K$, assuming $K>M$
    \begin{align}
  \frac{1}{K}\sum_k c_{k,m} c_{k,m'}= \begin{cases}1 & \text { if } m=m' \\ 0 & \text { otherwise }\end{cases}.
    \end{align}
    Other choices include Chu sequence, optimized binary phase codes, and phase codes that spread the inter-antenna interference in the Doppler domain~\cite{FegerHaderer16, SunPetropulu20}. 
    \item TDM-MIMO mode: the codes at Tx $m$ is a one-hot vector where $c_{k,m}=1$ and $c_{k,m'}=0, m' \neq m$. In other words, only $1$ Tx is active during one pulse and each Tx takes turns transmitting. 
    \item phased array mode: the codes at Tx $m$ are $1$, i.e., $c_{k,m}=1$ for all $k$. The beamforming angle is controlled by an additional beamforming process which is omitted here.
\end{itemize}

\subsection{Object Signal Model} \label{sec: Target}

Following the receiver processing at the victim radar of Fig.~\ref{fig:mimoBlockDiagram}, 
we provide a quick overview of object signal model in the spatial domain, e.g., Tx transmitting and Rx receiving angles. Similar derivation of the object signal model can be found in~\cite{WangBoufounos20,MIMOmodel}.

For an object of range $R$ and relative radial velocity $v$, the round-trip propagation delay from victim radar's $m$-th Tx antenna to its $n$-th receiving antenna is
\begin{align}
    \tau_{m,n} (t) = 2\frac{R + v t}{c} + m \frac{d_t \sin(\phi_{t})}{c} + n \frac{d_r \sin(\phi_{r})}{c},
\end{align}
where $d_t$ and $d_r$ are the Tx and Rx antenna element spacing, $\phi_{t}$ and $\phi_{r}$ are the Tx and Rx angle for the object, and $c$ is the speed of propagation. If the object is in the far field, we have the approximation $\phi_{t} = \phi_{r}$.

As shown in the upper right (victim Rx) of Fig.~\ref{fig:mimoBlockDiagram}, the received signal goes through processing blocks such as local oscillator (LO), low-pass filtering (LPF), analog-to-digital converter (ADC), fast-time/range FFT, slow-time/Doppler FFT, and MIMO waveform separation at each receiver antenna chain. A step-by-step derivation of the object signal model is included in Appendix~\ref{app: object}. At the output of the MIMO waveform separation and by stacking $\{y^s_{m,n}(l',k')\}$ into a vector, one can form an $MN \times 1$ virtual array signal for an object at a given pair of range bin $l'$ and Doppler bin $k'$ as
\begin{align} \label{ys}
    \mathbf y^s(l',k') = b(l',k') \mathbf a_{t} \otimes  \mathbf a_{r}.
\end{align}
where $\mathbf a_{t} \triangleq [1, e^{-j2\pi f_{\phi_t}}, \ldots, e^{-j2\pi f_{\phi_t}(M-1)}]^T$ 
is the object Tx steering vector with a spatial frequency of $f_{\phi_t}$, 
$\mathbf a_{r} \triangleq [1, e^{-j2\pi f_{\phi_r}}, \ldots, e^{-j2\pi f_{\phi_r} (N-1)}]^T$ is the object Rx steering vector with a spatial frequency of $f_{\phi_r}$, and $\otimes$ represents the Kronecker product.
It is seen that the spatial-domain object signal has a Kronecker structure between the object Tx and Rx steering vectors.

\subsection{Interference Signal Model} \label{sec: Interference}
In the lower left of Fig.~\ref{fig:mimoBlockDiagram},  an interfering radar also employs the MIMO-FMCW signaling scheme with possibly different MIMO array configurations such as the number of Tx antennas $\widetilde M$, inter-element spacing, and Tx-pulse codes $\widetilde c_{\widetilde k,\widetilde m}$, FMCW configuration parameters, and time offsets. 

\emph{Transmitted MIMO-FMCW Waveform at Interfering TX}: More specifically, the $\widetilde m$-th interfering Tx antenna sends coded  $\widetilde K$ pulses
\begin{align}
    \widetilde s_{\widetilde m}(t) = \sum_{\widetilde k=0}^{\widetilde K-1} \widetilde c_{\widetilde k,\widetilde m} \widetilde s(t-\widetilde k \widetilde T_{\text{PRI}} - \widetilde \tau_{syn}) e^{j2\pi f_c (t -\widetilde k \widetilde T_{\text{PRI}} - \widetilde \tau_{syn})},
\end{align}
where the source FMCM waveform $\widetilde s(t)$ shares the same expression as \eqref{eqn: FMCW} but with different chirp rate $\widetilde \beta$ and pulse duration $\widetilde T$, $\widetilde \tau_{syn}$ is the transmit synchronization delay (initial time offset) between the reference Tx antennas of the victim radar and the interfering radar, $\widetilde c_{\widetilde k,\widetilde m}$ is the Tx-pulse code of the interfering radar that likely are different from those used at the victim Tx), and $\widetilde T_{\text{PRI}}$ is the PRI at the interfering radar. 

\emph{Interference Waveform at Receiving Antennas of Victim Rx}: 
For an interfering radar of range $\widetilde R$ and radial velocity $\widetilde v$ relative to the victim radar, the one-way propagation delay from its $\widetilde m$-th Tx antenna to the $n$-th Rx antenna of victim radar is
\begin{align}
    \widetilde \tau_{\widetilde m,n} (t) = \frac{\widetilde R + \widetilde v t}{c} + \widetilde m \frac{\widetilde d_t \sin(\widetilde \phi_{t})}{c} + n \frac{d_r \sin(\widetilde \phi_{r})}{c},
\end{align}
where $\widetilde d_t$ is the Tx antenna element spacing at the interferer, and $\widetilde \phi_{t}$ and $\widetilde \phi_{r}$ are the interference Tx and Rx angles with respect to the boresight of the interfering radar and the victim radar. At the victim Rx of Fig.~\ref{fig:mimoBlockDiagram}, the $n$-th receiver observes the RF signal from the interferer
\begin{align}
   & s^i_n(t) = \widetilde \alpha \sum_{\widetilde m=0}^{\widetilde M-1} \widetilde s_{\widetilde m}(t - \widetilde \tau_{\widetilde m,n} (t)) \nonumber\\
   \approx & \widetilde \alpha e^{- j2\pi f_c \widetilde \tau} \sum_{\widetilde m=0}^{\widetilde M-1} \sum_{\widetilde k=0}^{\widetilde K-1} \widetilde c_{\widetilde k,\widetilde m} \widetilde s(t-\widetilde k \widetilde T_{\text{PRI}} - \widetilde \tau_{syn} - \widetilde \tau)   \nonumber\\
   & \times e^{j2\pi f_c (t-\widetilde k \widetilde T_{\text{PRI}}-\widetilde \tau_{syn})} e^{- j2\pi (\widetilde f_{\phi_t} \widetilde m + \widetilde f_{\phi_r} n)}  e^{-j2\pi f_c \frac{\widetilde vt}{c}},
\end{align}
where $\widetilde \alpha$ is the received complex amplitude of the interference, $\widetilde \tau = {\widetilde R}/{c}$ is the reference one-way propagation delay from interferer to the victim radar, and $\widetilde f_{\phi_t} = \widetilde d_t {\sin(\widetilde \phi_{t})}/{\lambda}$ and $\widetilde f_{\phi_r} = d_r {\sin(\widetilde \phi_{r})}/{\lambda}$ are the normalized spatial frequency at the interferer transmitting antennas and victim receiving antennas.

\emph{Interference Waveform after LO and LPF at Victim Rx}: Since our goal is to derive the interference signal model seen at the victim Rx, we need to convert the interference time, i.e., $\widetilde k \widetilde T_{\text{PRI}} + \widetilde \tau_{syn}, \widetilde k = 0,1,\ldots, \widetilde K-1$ to the reference of the victim radar. As details are shown in Appendix~\ref{app: interference} and defining $\widetilde \tau'_{k,\widetilde k}$ as the time offset between the $\widetilde k$-th pulse of the interfering radar relative to the $k$-th pulse at the victim radar, we can express the low-pass filtered interference signal $a^{s,low}_n(t)$ sampled at $t = kT_{\text{PRI}} + l \Delta T$ as 
\begin{align}
    & a^{i}_n(l,k)  = a^{i,low}_n(kT_{\text{PRI}} + l \Delta T) \nonumber\\
   = & \widetilde \alpha e^{- j2\pi f_c \widetilde \tau} \sum_{\widetilde m=0}^{\widetilde M-1} \sum_{\widetilde k=0}^{\widetilde K-1} \widetilde c_{k,\widetilde m}^{\widetilde k} e^{j\pi \widetilde \beta (\widetilde \tau'_{k,\widetilde k} + \widetilde \tau)^2} e^{-j2\pi f_c  \widetilde \tau'_{k,\widetilde k}} \nonumber\\
   & \times  e^{j\pi (\widetilde \beta - \beta) (l \Delta T)^2}  
    e^{-j2\pi \widetilde f_{r,k,\widetilde k} l} \mathbf{1} \left[l \in \mathcal L^i_{k,\widetilde k}\right] \nonumber\\
   & \times e^{-j2\pi \widetilde f_d k} e^{- j2\pi (\widetilde f_{\phi_t} \widetilde m + \widetilde f_{\phi_r} n)}  
\end{align}
where $\widetilde c_{k,\widetilde m}^{\widetilde k}$ is the  slow-time code of the interfering radar observed at $k$-th victim radar pulse and is defined in Appendix~\ref{app: interference}, $\mathbf{1}[\cdot]$ is the indicator function, 
\begin{align}
\mathcal L^i_{k,\widetilde k} \triangleq \Big\{l: & 0<  \widetilde \beta (\widetilde \tau'_{k,\widetilde k} + \widetilde \tau) - (\widetilde \beta - \beta) l \Delta T < f_L, \\
& (\widetilde \tau'_{k,\widetilde k} + \widetilde \tau) < l\Delta T < \min\left\{T, \widetilde \tau'_{k,\widetilde k} + \widetilde \tau + \widetilde T\right\}\Big\} \notag 
\end{align}
is the set of interference contaminated sample indices, $\widetilde f_{r,k,\widetilde k} \triangleq (\widetilde \beta(\widetilde \tau'_{k,\widetilde k} + \widetilde \tau) + \frac{\widetilde v}{\lambda})\Delta T$ is the normalized interference initial fast-time frequency, and $\widetilde f_d = f_c {\widetilde v T_{\text{PRI}}}/{c}$ is the normalized interference Doppler frequency.

\emph{Interference Waveform after range FFT at Victim Rx}:
Applying the range FFT to the sampled interference signal $a^{i}_n(l,k)$, we obtain its range spectrum at the $n$-th Rx antenna, $l'$-th range bin and $k$-th pulse as 
\begin{align} 
    & x^i_{n}(l',k) 
   = \sum_{\widetilde m=0}^{\widetilde M-1} \widetilde \alpha_{l',k,\widetilde m} e^{-j2\pi \widetilde f_d k} e^{- j2\pi (\widetilde f_{\phi_t} \widetilde m + \widetilde f_{\phi_r} n)},
\end{align}
where 
\begin{align} \label{eqn: intCodedCompAmp}
     &\widetilde \alpha_{l',k,\widetilde m} \triangleq  \widetilde \alpha e^{- j2\pi f_c \widetilde \tau} \sum_{\widetilde k=0}^{\widetilde K-1} \widetilde c_{k,\widetilde m}^{\widetilde k} e^{j\pi \widetilde \beta (\widetilde \tau'_{k,\widetilde k} + \widetilde \tau)^2} e^{-j2\pi f_c  \widetilde \tau'_{k,\widetilde k}} \nonumber\\
    &\times  \sum_{l=0}^{L-1} e^{j\pi (\widetilde \beta - \beta) (l \Delta T)^2}  \mathbf{1} \left[l \in \mathcal L^i_{k,\widetilde k}\right] e^{-j2\pi (\widetilde f_{r,k,\widetilde k} +\frac{l'}{L}) l}
\end{align}
is the coded complex interference amplitude from the $\widetilde m$-th interfering Tx antenna at victim radar's range bin $l'$ and pulse $k$.
It is worthy noting that $\widetilde \alpha_{l',k,\widetilde m}$ varies with pulse index $k$ since it is a function of $\widetilde \tau'_{k,\widetilde k}$. Moreover, $\widetilde \alpha_{l',k,\widetilde m}$ is unknown at the victim Rx. 

\emph{Interference Waveform after Doppler FFT and Waveform Separation at Victim Rx}:
For MIMO waveform separation, the victim Rx only applies the same procedure using the Tx-pulse code from the victim Tx and assumes no prior knowledge about the Tx-pulse code of the interfering Tx. As a result, the interference spectrum at $l'$-th range bin and $k'$-th Doppler bin of the victim Rx is given as 
\begin{align} \label{eqn: intRDSpec}
    & y^i_{m,n}(l',k') = \sum_{k=0}^{K-1} x^i_{n}(l',k) c_{k,m}^* e^{-j2\pi \frac{k'}{K}k} \nonumber\\
    = & \sum_{\widetilde m=0}^{\widetilde M-1} \sum_{k=0}^{K-1} \widetilde \alpha_{l',k,\widetilde m} c_{k,m}^*  e^{-j2\pi (\widetilde f_d + \frac{k'}{K}) k} e^{-j2\pi (\widetilde f_{\phi_t} \widetilde m + \widetilde f_{\phi_r} n)} \nonumber \\
    = & \widetilde a'_{t,m} e^{-j2\pi \widetilde f_{\phi_r} n},
\end{align}
where the Tx-pulse codes $c_{k,m}$ used at the victim Tx are used for the waveform separation, 
\begin{align}\label{eqn: decodedIntTxSteer}
    \widetilde a'_{t,m} = \sum_{\widetilde m=0}^{\widetilde M-1} \sum_{k=0}^{K-1} \widetilde \alpha_{l',k,\widetilde m} c_{k,m}^*  e^{-j2\pi (\widetilde f_d + \frac{k'}{K}) k} e^{-j2\pi \widetilde f_{\phi_t} \widetilde m}.
\end{align}

\emph{Spatial-Domain Interference Steering Vector at Victim Rx}:
Stacking $\{y^i_{m,n}(l',k')\}$ into a vector, we obtain the interference range-Doppler spectrum on an $MN \times 1$ virtual array
\begin{align} \label{eqn: decodedIntVirtual}
    \mathbf y^i(l',k') = \widetilde {\mathbf a}'_{t} \otimes  \widetilde{\mathbf a}_{r}.
\end{align}
where 
\begin{align} \label{eqn: decodedIntTxSteerVec}
    \widetilde {\mathbf a}'_{t} \triangleq [\widetilde a'_{t,0}, \widetilde a'_{t,1}, \ldots, \widetilde a'_{t,M-1}]^T, 
\end{align}
is the $M \times 1$ interfering Tx steering signal seen at the victim Rx, and
\begin{align}  \label{inteferingRxVec}
    \widetilde{\mathbf a}_{r} \triangleq [1, e^{-j2\pi \widetilde f_{\phi_r}}, \ldots, e^{-j2\pi \widetilde f_{\phi_r} (N-1)}]^T 
\end{align}
is the $N \times 1$ interfering Rx steering vector. 

From \eqref{eqn: decodedIntVirtual}, it is seen that the spatial-domain interference steering vector also has the Kronecker structure between the Tx and Rx steering vectors, like the spatial-domain object steering vector in \eqref{ys}. The main difference lies in the interference Tx steering vector of \eqref{eqn: decodedIntTxSteer} which is a function of the transmitting power of the interfering radar, interfering-victim relative distance and Doppler frequency, FMCW time-frequency incoherence (e.g., chirp rate, pulse duration, pulse repetition interval), MIMO incoherence (e.g., MIMO code and Tx array configuration), and timing offset between the interfering and victim radars. In other words, the object Tx/Rx steering vectors and interfering Rx steering vector are fully determined by the object-victim and interfering-victim directions due to their Fourier vector structure, while the interfering Tx steering vector is almost unknown as its direction in the $M$-dimensional subspace is not only determined by the relative interfering-victim direction but also the mentioned incoherence.

\subsection{Examples of MIMO-FMCW Interference Signal Model}

In the following, we show that our derived MIMO-FMCW interference model reduces to three special interference scenarios widely used in the existing literature when certain conditions are met. 

\subsubsection{Coherent Interference}

In this part, we validate that when interferer and victim radar are synchronized ($\widetilde \tau_{syn} = 0$), have the same waveform parameters ($\widetilde \beta = \beta$, $\widetilde T_{\text{PRI}} = T_{\text{PRI}}$, $\widetilde T = T$, $\widetilde K = K$), number of Tx antennas ($\widetilde M = M$) and slow-time code ($\{\widetilde c_{\widetilde k,\widetilde m}\} = \{c_{k,m}\}$), the received interference signal, referred to as the coherent interference signal, has the same structure as the object signal~\cite{SianJSTSP}.

Under the coherent interference step, by \eqref{eqn: IntKset}, we have $\tau'_{k,\widetilde k} = 0$ and $\mathcal K_{\widetilde k} = \{\widetilde k\}$.
Then, by \eqref{eqn: rxIntMIMOcode}, we have $\widetilde c_{k,\widetilde m}^{\widetilde k} = \widetilde c_{k,\widetilde m} = c_{k,\widetilde m}$ if $k = \widetilde k$, and otherwise $\widetilde c_{k,\widetilde m}^{\widetilde k} = 0$; as here we consider the coherent interference is dechirped into the victim radar, i.e., $0<  \widetilde \beta \widetilde \tau < f_L$, we have $\mathcal L^i_{k,\widetilde k} = \{\lceil {\widetilde \tau}/{\Delta T} \rceil,\ldots, \lfloor {T}/{\Delta T}\rfloor\}$; the normalized interference initial fast-time frequency reduces to $\widetilde f_{r,k,\widetilde k} = (\widetilde \beta \widetilde \tau + \frac{\widetilde v}{\lambda})\Delta T$.
Based on these $\widetilde \alpha_{l',k,\widetilde m}$ in \eqref{eqn: intCodedCompAmp} becomes $\widetilde \alpha_{l',k,\widetilde m} = \widetilde \alpha_{l'} c_{k,\widetilde m}$, where $\widetilde \alpha_{l'} = \widetilde \alpha e^{- j2\pi f_c \widetilde \tau} e^{j\pi \widetilde \beta \widetilde \tau^2} \sum_{l=0}^{L-1} \mathbf{1} \left[l \in \mathcal L^i_{k,\widetilde k}\right] e^{-j2\pi (\widetilde f_{r,k,\widetilde k} +\frac{l'}{L}) l}$.
Then, the range-Doppler interference spectrum in \eqref{eqn: intRDSpec} reduces to 
\begin{align} \label{eqn: intRDSpec2}
    & y^i_{m,n}(l',k') \nonumber\\
    = & \widetilde \alpha_{l'} \sum_{\widetilde m \neq m} \left(\sum_{k=0}^{K-1} c_{k,\widetilde m}  c_{k,m}^* e^{-j2\pi (\widetilde f_d + \frac{k'}{K}) k} \right) e^{-j2\pi (\widetilde f_{\phi_t} \widetilde m + \widetilde f_{\phi_r} n)}  \nonumber\\
     & + b^i(l',k') e^{-j2\pi (\widetilde f_{\phi_t} m + \widetilde f_{\phi_r} n)},
\end{align}
where $b^i(l',k') \triangleq \widetilde \alpha_{l'} \left(\sum_{k=0}^{K-1}  e^{-j2\pi (\widetilde f_d + \frac{k'}{K}) k}\right)$.
Notice that when $\widetilde f_d + \frac{k'}{K} \approx 0$, i.e., when the normalized interference Doppler frequency fall near the Doppler bin $k'$, then $b^i(l',k') \approx K \widetilde \alpha_{l'}$ indicating a peak on Doppler spectrum, and in this case
\begin{align} \label{eqn: intRDySignalTarDoppBin}
    y^i_{m,n}(l',k') \approx & b^i(l',k') e^{-j2\pi (\widetilde f_{\phi_t} m + \widetilde f_{\phi_r} n)}
\end{align}
due to \eqref{eqn: orthogonality}. Comparing the object signal $y^s_{m,n}(l',k')$ in \eqref{eqn: RDySignal} and the interference signal $y^i_{m,n}(l',k')$ in \eqref{eqn: intRDySignalTarDoppBin}, we validate that under the coherent interference case, the interference model derived in Section~\ref{sec: Interference} has a similar structure compared to the object signal model derived in Section~\ref{sec: Target}.

\subsubsection{Phased Array Radar Interference}

In this part, we show that when all radars are phased array radar~\cite{JianMIMOMSP} with slow-time code ($\{c_{k,m} = 1\}$), the spatial-domain interference signal has a Fourier structure.

Under the phased array radar setup, $\{c_{k,m} = 1\}$ implies that $\{\widetilde c_{k,\widetilde m}^{\widetilde k}=1\}$ and $\{c_{k,m}^* = 1\}$.
Then, $\widetilde \alpha_{l',k,\widetilde m}$ in~\eqref{eqn: intCodedCompAmp} is independent of $\widetilde m$ and $y^i_{m,n}(l',k')$ in~\eqref{eqn: intRDSpec} is independent of $m$.
Rewriting $\widetilde \alpha_{l',k,\widetilde m}$ as $\widetilde \alpha_{l',k}$ and rewriting $y^i_{m,n}(l',k')$ as $y^i_{n}(l',k')$, the range-Doppler interference spectrum in \eqref{eqn: intRDSpec} reduces to 
\begin{align}
    y^i_{n}(l',k') = \widetilde a'_{t} e^{-j2\pi \widetilde f_{\phi_r} n},
\end{align}
where
\begin{align}
    \widetilde a'_{t} = \sum_{k=0}^{K-1} \widetilde \alpha_{l',k} e^{-j2\pi (\widetilde f_d + \frac{k'}{K}) k} \left(\sum_{\widetilde m=0}^{\widetilde M-1} \widetilde w_{\widetilde m} e^{-j2\pi \widetilde f_{\phi_t} \widetilde m}\right),
\end{align}
and $\widetilde w_{\widetilde m}$ is the Tx beamforming weights on $\widetilde m$-th interference Tx antenna.
Stacking $\{y^i_{n}(l',k')\}$ into a vector, we obtain the interference range-Doppler spectrum on a $N \times 1$ Rx array
\begin{align} \label{eqn: decodedIntPhased}
    \mathbf y^i(l',k') = \widetilde a'_{t} \widetilde{\mathbf a}_{r},
\end{align}
which is a Fourier vector. 
Notice that this interference structure also applies in the special case where all radars adopt a single Tx antenna.

\subsubsection{TDM-MIMO Radar Interference}

In this part, we show that when all radars are TDM-MIMO radars, the spatial-domain interference signal has the same structure as in \eqref{eqn: decodedIntVirtual}.

Mathematically, the modification in the above derivation is in two folds. First, the slow-time phase code $c_{k,m}$ is replaced as the slow-time code with $c_{k,m} = 1$ if $m = \mod (k,M)$ and $c_{k,m} = 0$ otherwise, for $ k = 0,1,\ldots, K-1$ and $m = 0,1,\ldots, M-1$. Second, in the Doppler FFT equations \eqref{eqn: RDySignal} and \eqref{eqn: intRDSpec}, $e^{-j2\pi \frac{k'}{K}k}$ is replaced by $e^{-j2\pi \frac{k'}{\lfloor K/M \rfloor}k}$ because only $\lfloor K/M \rfloor$ pulses are used in TDM-MIMO for each antenna.
These two modifications do not affect the interference structure in \eqref{eqn: decodedIntVirtual}.

\section{Problem Formulation}
In this section, we first formulate object detection as a composite hypothesis testing problem and review existing detectors. 

\subsection{Spatial-domain Detection Problem under Interference}
Given the target and interference signal models over a given range-Doppler bin, the spatial-domain object detection under mutual interference is formulated as a composite hypothesis testing problem 
\begin{align}  \label{eqn: IntHyp}
\begin{cases}
    H_0, \ & \mathbf y = \sum_{q=1}^Q \widetilde {\mathbf a}'_{t,q} \otimes \widetilde{\mathbf a}_{r,q} + \mathbf z \\
    H_1, \ & \mathbf y =  b \mathbf a_{t} \otimes  \mathbf a_{r} + \sum_{q=1}^Q \widetilde {\mathbf a}'_{t,q} \otimes \widetilde{\mathbf a}_{r,q} + \mathbf z,
\end{cases}
\end{align}
where $\mathbf y$ is the complex-valued range-Doppler spectrum at a given range-Doppler bin $(l', k')$, $b$ is the complex-valued unknown object amplitude, $Q$ is the number of interference, $\mathbf{a}_t$ and $\mathbf{a}_r$ are object Tx and Rx steering vectors defined below~\eqref{ys}, $\widetilde{\mathbf a}'_{t,q}$ and $\widetilde{\mathbf a}_{r,q}$ are the $q$-th decoded interference Tx and Rx steering vector given in the form of~\eqref{eqn: decodedIntTxSteerVec} and~\eqref{inteferingRxVec}, and the noise $\mathbf z \sim \mathcal {CN}(\mathbf{0}, \sigma^2 \mathbf I_{MN})$ with $\mathbf I_{MN}$ denoting the identity matrix of size $MN$. The null hypothesis $H_0$ consists of interference and noise, and the alternative hypothesis $H_1$ consists of the object signal plus interference and noise.

It is worth noting that, in \eqref{eqn: IntHyp}, we assume the knowledge of the interference Rx steering vector $\widetilde{\mathbf a}_{r}$. This assumption on $\widetilde{\mathbf a}_{r,q}$ is motivated by the observation that it is a Fourier vector at the angle of the interfering radar. 
We assume the angle of arrival in $\widetilde{\mathbf a}_{r,q}$ and the number of interference $Q$ can be estimated from nearby range-Doppler bins.

\subsection{Existing Detectors}

\subsubsection{Clairvoyant Detector}
Assuming the perfect knowledge of $\widetilde{\mathbf a}'_{t}$, the clairvoyant detector is given by
\begin{align} \label{eqn: clairvoyant}
    & T^C(\mathbf y)  = \frac{2}{\sigma^2} \frac{\left| (\mathbf a_{t} \otimes  \mathbf a_{r})^H (\mathbf{y}-\sum_{q=1}^Q \widetilde {\mathbf a}'_{t,q} \otimes \widetilde{\mathbf a}_{r,q}) \right|^2}{||\mathbf a_{t} \otimes  \mathbf a_{r}||^2}. 
\end{align}
It is equivalent to removing all interference components $\sum_{q=1}^Q \widetilde {\mathbf a}'_{t,q} \otimes \widetilde{\mathbf a}_{r,q}$ before the matched filtering with respect to the object steering vector.
The probabilities of false alarm and detection of \eqref{eqn: clairvoyant} can be derived as
\begin{align} \label{pfpd_clairvoyant}
    P_{FA}^{C} = e^{-\frac{1}{2}\gamma}, \quad P_{D}^{C} = Q_1 \left(\sqrt{\lambda^C},\sqrt{\gamma}\right), 
\end{align}
where $\gamma$ is the threshold used for detection, and the parameter $\lambda^c$ is given as
\begin{align}
\lambda^C = \frac{2|b|^2}{\sigma^2} \left|\left|\mathbf a_{t} \otimes \mathbf a_{r} \right|\right|^2 = \frac{2MN|b|^2}{\sigma^2}.
\end{align}
It is worthy noting that the clairvoyant detector of \eqref{eqn: clairvoyant} cannot be implemented in practice due to the strong assumption about the incoherent interference Tx steering vector $\widetilde{\mathbf a}'_{t}$. 

\subsubsection{Receiver Subspace (RS) Detector of \cite{SAM22}} Assuming the number of Tx antennas is small, one can directly treat $\widetilde{\mathbf a}'_{t}$ as an nuisance parameter in~\eqref{eqn: IntHyp} and estimate it under both hypotheses. The resulting generalized likelihood ratio test (GLRT) is given by \cite{SAM22}
\begin{align} \label{eqn: RS}
    & T^{RS}(\mathbf y) = \frac{2}{\sigma^2} \frac{\left|\left(\mathbf a_{t} \otimes \left(\mathbf P_{\widetilde{\mathbf A}_{r}}^{\perp} \mathbf a_{r}\right)\right)^H\mathbf y\right|^2}{\left|\left|\mathbf a_{t} \otimes \left(\mathbf P_{\widetilde{\mathbf A}_{r}}^{\perp} \mathbf a_{r}\right) \right|\right|^2}, 
\end{align}
where $\widetilde{\mathbf A}_{r} \triangleq \left[\widetilde{\mathbf a}_{r,1}, \widetilde{\mathbf a}_{r,2}, \ldots, \widetilde{\mathbf a}_{r,Q}\right]$ is a stack of $Q$ interference Rx steering vectors. It is clear to see that the GLRT only exploits the receiver subspace $\widetilde{\mathbf A}_{r}$ of the interferences. Correspondingly, 
the probabilities of false alarm probability and detection are given by
\begin{align} \label{pfpd_rs}
    P_{FA}^{RS}= e^{-\frac{1}{2}\gamma}, \quad P_{D}^{RS}= Q_1 \left(\sqrt{\lambda^{RS}},\sqrt{\gamma}\right),
\end{align}
where 
\begin{align}
\lambda^{RS}= \frac{2|b|^2}{\sigma^2} \left|\left|\mathbf a_{t} \otimes \left(\mathbf P_{\widetilde{\mathbf A}_{r}}^{\perp} \mathbf a_{r}\right) \right|\right|^2.
\end{align}

\subsubsection{LCMV Detector of \cite{EuRADMIMO}}
In~\cite{EuRADMIMO}, a conventional linear constraint minimum variance (LCMV) beamformer is adopted. It assumes that
\begin{align} \label{eqn:IplusN}
    \sum_{q=1}^Q \widetilde {\mathbf a}'_{t,q} \otimes \widetilde{\mathbf a}_{r,q} + \mathbf z \sim \mathcal {CN}(\mathbf 0, \sigma^2 \widetilde {\mathbf R}),
\end{align}
where $\widetilde {\mathbf R}$ is a normalized covariance matrix. Assuming the perfect knowledge of $\widetilde {\mathbf R}$, the LCMV solves the following optimization problem \cite{van2004optimum}:
\begin{align} \label{eqn: LCMVBFProb}
    \min_{\mathbf w} \; \; & \mathbf w^H \widetilde {\mathbf R} \mathbf w \nonumber\\
    s.t.  \;\;  & (\mathbf a_{t} \otimes  \mathbf a_{r})^H\mathbf w = 1.
\end{align}
which leads to the LCMV detector \cite{EuRADMIMO}:
\begin{align} \label{eqn: TLCMV}
    T^{LCMV}(\mathbf y) = \frac{2}{\sigma^2} \frac{\left| \left(\widetilde {\mathbf R}^{-1}  (\mathbf a_{t} \otimes  \mathbf a_{r})\right)^H\mathbf y\right|^2}{\Big|\Big|\widetilde {\mathbf R}^{-\frac{1}{2}} (\mathbf a_{t} \otimes  \mathbf a_{r})\Big|\Big|^2}.
\end{align}
Given the knowledge of $\widetilde {\mathbf R}$,  the probabilities of false alarm  and detection are given by
\begin{align} 
    P_{FA}^{LCMV} = e^{-\frac{1}{2}\gamma}, \quad P_{D}^{LCMV} = Q_1 \left(\sqrt{\lambda^{LCMV}},\sqrt{\gamma}\right),
\end{align}
where 
\begin{align}
\lambda^{LCMV} = \frac{2|b|^2}{\sigma^2} \left\|\widetilde {\mathbf R}^{-\frac{1}{2}} (\mathbf a_{t} \otimes  \mathbf a_{r})\right\|^2.
\end{align}

\section{Proposed Object Detection under MIMO-FMCW Mutual Interference}
\label{sec: GS}

In the following, we first demonstrate limitations inherent in the RS and LCMV detectors and gain insights through an in-depth examination of the clairvoyant detector. Then, we propose a  generalized subspace (GS) detector that leverages both the Tx and Rx steering vectors of the interference, followed by a comprehensive theoretical performance analysis of the detection performance under the mutual interference. 

\subsection{Observations from Existing Detectors}
For the RS detector of \eqref{eqn: RS}, it projects each interference signal $\widetilde {\mathbf a}'_{t,q} \otimes \widetilde{\mathbf a}_{r,q}, q = 1,2,\ldots, Q$ to $0$, i.e., 
\begin{align} \label{eqn: orthogonalInt}
    (\mathbf a_{t} \otimes (\mathbf P_{\widetilde{\mathbf A}_{r}}^{\perp} \mathbf a_{r}))^H(\widetilde {\mathbf a}'_{t,q} \otimes \widetilde{\mathbf a}_{r,q})  = 0,
\end{align}
because the interference Rx steering vector $\widetilde{\mathbf a}_{r,q}$ is projected to its orthogonal subspace, i.e., $(\mathbf P_{\widetilde{\mathbf A}_{r}}^{\perp} \mathbf a_{r})^H\widetilde{\mathbf a}_{r,q} = 0$. However, this operation fails to maintain the matched filtering gain with respect to the object steering vector as
\begin{align}
    (\mathbf a_{t} \otimes (\mathbf P_{\widetilde{\mathbf A}_{r}}^{\perp} \mathbf a_{r}))^H (\mathbf a_{t} \otimes  \mathbf a_{r}) = M \mathbf a_{r}^H\mathbf P_{\widetilde{\mathbf A}_{r}}^{\perp} \mathbf a_{r} <MN,
\end{align}
where $MN$ is the coherent matched filtering gain.
This is undesirable, particularly when the interference power is small, as the RS detector may mitigate low-power interference at the price of losing object detection gain.

For the LCMV detector of \eqref{eqn: TLCMV}, it is sensitive to adaptive estimation error of $\widetilde {\mathbf R}$ when $\widetilde {\mathbf R}$ is not known \emph{a priori}.  Obtaining an accurate estimate of $\widetilde {\mathbf R}$ requires a large number of homogeneous and object-free training samples, which may be challenging in the presence of dense automotive radars. 
 
\begin{figure}[t]
\begin{center}
  {\resizebox{4.5cm}{!}{\includegraphics{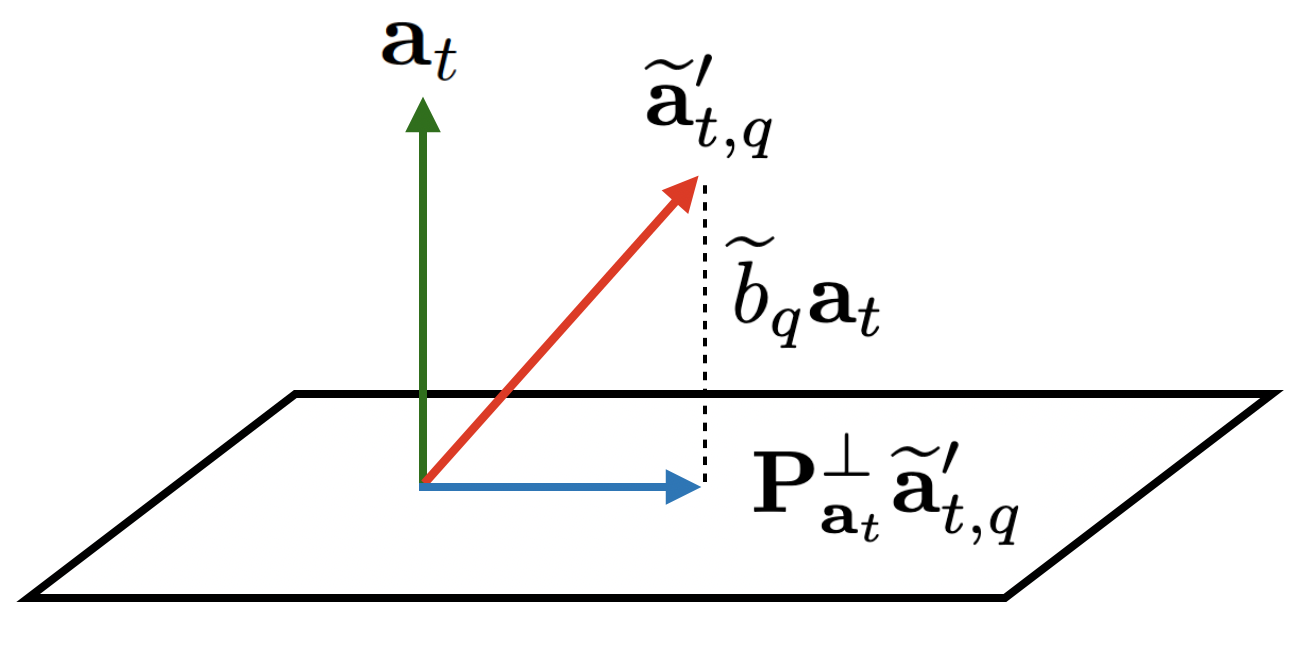}}}
         \caption{\small{Decomposition of ${\widetilde {\mathbf a}'}_{t,q}$ into $\widetilde b_q \mathbf a_{t}$ and $\mathbf P_{\mathbf a_{t}}^{\perp} {\widetilde {\mathbf a}'}_{t,q}$ in a 3-D example, where the plane is the orthogonal subspace of $\mathbf a_{t}$.
          }}\label{fig:decomp}
\end{center}
\vspace{-0.22in}
\end{figure}

Finally, for the clairvoyant detector of \eqref{eqn: clairvoyant}, one can decompose the $q$-th interference Tx steering vector along with the object Tx steering vector and its orthogonal complement direction
\begin{align} \label{eqn: decom}
    {\widetilde {\mathbf a}'}_{t,q} = \widetilde b_q \mathbf a_{t} + \mathbf P_{\mathbf a_{t}}^{\perp} {\widetilde {\mathbf a}'}_{t,q}
\end{align}
as shown in Fig.~\ref{fig:decomp}, where the resulting amplitude along $\mathbf a_{t}$ is given
\begin{align} \label{eqn: btuta}
    \widetilde b_q = \frac{\mathbf a_{t}^H{\widetilde {\mathbf a}'}_{t,q}}{||\mathbf a_{t}||^2}.
\end{align}
With \eqref{eqn: decom},  the clairvoyant detector of \eqref{eqn: clairvoyant} can be rewritten as
\begin{align} \label{eqn: clairvoyant2}
    & T^{C}(\mathbf y) = \frac{2}{\sigma^2} \frac{\left| (\mathbf a_{t} \otimes  \mathbf a_{r})^H (\mathbf{y}-\sum_{q=1}^Q\widetilde b_q \mathbf a_{t} \otimes  \widetilde{\mathbf a}_{r,q})\right|^2}{||\mathbf a_{t} \otimes  \mathbf a_{r}||^2},
\end{align}
which implies that the essential interference to cancel is a rank-$Q$ interference with known directions $\mathbf a_{t} \otimes  \widetilde{\mathbf a}_{r,q}, q = 1,2,\ldots, Q$, and the unknown parameters sufficient for interference cancellation is $\widetilde b_q, q = 1,2,\ldots, Q$.

\subsection{Proposed Generalized Subspace (GS) Detector}
Since the exact knowledge of $\widetilde b_q$ is unknown in advance, we may estimate the power of $\widetilde b_q$, denoted by $h_q^2$, from nearby range-Doppler bins.
Assume $\widetilde b_q \sim \mathcal {CN}(0, h_q^2)$ with known variance $h_q^2$ and $\widetilde b_q$ is independent of $\mathbf z$.
Then, the essential interference plus noise is 
\begin{align} \label{eqn: z_tu_ta}
     \widetilde {\mathbf z} = \sum_{q=1}^Q\widetilde b_q \mathbf a_{t} \otimes  \widetilde{\mathbf a}_{r,q} + \mathbf z \sim \mathcal {CN}(\mathbf 0, \sigma^2 \mathbf R),
\end{align}
and the normalized covariance of $\widetilde {\mathbf z}$ is
\begin{align} \label{eqn: R_cov}
    & \mathbf R =  \sum_{q=1}^Q \frac{h_q^2}{\sigma^2} (\mathbf a_{t} \otimes  \widetilde{\mathbf a}_{r,q}) (\mathbf a_{t} \otimes  \widetilde{\mathbf a}_{r,q})^H + \mathbf I_{MN}.
\end{align}
We first design our Rx beamformer $\mathbf w$ to satisfy the following criterion: 
\begin{enumerate}
    \item minimize the variance of interference-plus-noise with known covariance after beamforming, i.e., $\mathbf w^H \mathbf R \mathbf w$;
    \item maintain a fixed gain at the object direction, i.e.,  $(\mathbf a_{t} \otimes  \mathbf a_{r})^H\mathbf w = 1$; 
    \item force the the unknown interference $\sum_{q=1}^Q(\mathbf P_{\mathbf a_{t}}^{\perp} {\widetilde {\mathbf a}'}_{t,q}) \otimes  \widetilde{\mathbf a}_{r,q}$ to zero for any ${\widetilde {\mathbf a}'}_{t}$, i.e., 
    \begin{align}
    \sum_{q=1}^Q((\mathbf P_{\mathbf a_{t}}^{\perp} {\widetilde {\mathbf a}'}_{t,q}) \otimes  \widetilde{\mathbf a}_{r,q})^H\mathbf w = 0,
    \end{align}
    for any ${\widetilde {\mathbf a}'}_{t,q}, q=1,2,\ldots,Q$, which is equivalent to force $(\mathbf P_{\mathbf a_{t}}^{\perp} \otimes  \widetilde{\mathbf a}_{r,q})^H\mathbf w = \mathbf 0_{M}, q=1,2,\ldots,Q$, where $\mathbf 0_{M}$ denotes the $M$-dimensional column vector with all $0$ elements.
\end{enumerate}
As a result, one needs to solve the following optimization problem:
\begin{align} \label{eqn: GSRxBFProb}
    \min_{\mathbf w} \; \; & \mathbf w^H \mathbf R \mathbf w \nonumber\\
    s.t.  \;\;  & (\mathbf a_{t} \otimes  \mathbf a_{r})^H\mathbf w = 1, \nonumber\\
    & (\mathbf P_{\mathbf a_{t}}^{\perp} \otimes  \widetilde{\mathbf a}_{r,q})^H\mathbf w = \mathbf 0_{M}, q=1,2,\ldots,Q.
\end{align}
Compared to the LCMV beamforming optimization problem in~\eqref{eqn: LCMVBFProb}, the objective function of the problem in~\eqref{eqn: GSRxBFProb} is different in that it uses the essential interference plus noise covariance matrix $\mathbf R$ instead of the total interference plus noise covariance matrix $\widetilde {\mathbf R}$.

Denote $\widetilde{\mathbf A} \triangleq \left[\mathbf a_{t} \otimes  \widetilde{\mathbf a}_{r,1},\mathbf a_{t} \otimes  \widetilde{\mathbf a}_{r,2},\ldots, \mathbf a_{t} \otimes  \widetilde{\mathbf a}_{r,Q}\right]$ as the stack of $Q$ essential interference virtual steering vectors, and denote 
\begin{align} \label{eqn:lambda}
\boldsymbol \Lambda \triangleq \text{diag}\left[\frac{h_1^2}{\sigma^2},\frac{h_2^2}{\sigma^2},\ldots,\frac{h_Q^2}{\sigma^2}\right],
\end{align}
as the essential-interference-to-noise-ratio (EINR) matrix with diagonal elements reflecting the power values of $Q$ interferences over the noise. Then, we have the important observation
\begin{align} \label{eqn: R_inv}
    &\mathbf R^{-1} = \mathbf I_{MN} - \widetilde{\mathbf A} (\boldsymbol \Lambda^{-1} + \widetilde{\mathbf A}^H\widetilde{\mathbf A})^{-1}  \widetilde{\mathbf A}^H \nonumber\\
    = & \mathbf I_{MN} - \mathbf P_{\mathbf a_t} \otimes \widetilde{\mathbf P}_{\widetilde{\mathbf A}_r,\boldsymbol \Lambda},
\end{align}
where the regularized projection matrix 
\begin{align}
    \widetilde{\mathbf P}_{\widetilde{\mathbf A}_r,\boldsymbol \Lambda} = M\widetilde{\mathbf A}_r (\boldsymbol \Lambda^{-1} + M\widetilde{\mathbf A}_r^H\widetilde{\mathbf A}_r)^{-1}  \widetilde{\mathbf A}_r^H.
\end{align}
This special structure of $\mathbf R^{-1}$ implies that the optimal solution of the relaxed version of problem~\eqref{eqn: GSRxBFProb}
\begin{align} \label{eqn: GSRxBFProbRelax}
    \min_{\mathbf w} \; \; & \mathbf w^H \mathbf R \mathbf w \nonumber\\
    s.t.  \;\;  & (\mathbf a_{t} \otimes  \mathbf a_{r})^H\mathbf w = 1,
\end{align}
which is 
\begin{align} \label{eqn:w_GS}
    & \mathbf w^{GS} = \frac{\mathbf R^{-1}(\mathbf a_{t} \otimes  \mathbf a_{r})}{\Big|\Big| \mathbf R^{-\frac{1}{2}} (\mathbf a_{t} \otimes  \mathbf a_{r})\Big|\Big|^2} 
    = \frac{\mathbf a_{t} \otimes  (\widetilde{\mathbf P}_{\widetilde{\mathbf A}_r,\boldsymbol \Lambda}^{\perp}\mathbf a_{r})}{M \mathbf a_{r}^H \widetilde{\mathbf P}_{\widetilde{\mathbf A}_r,\boldsymbol \Lambda}^{\perp} \mathbf a_{r}},
\end{align}
satisfies the last condition of problem~\eqref{eqn: GSRxBFProb}, where $\widetilde{\mathbf P}_{\widetilde{\mathbf A}_r,\boldsymbol \Lambda}^{\perp} \triangleq \mathbf I_N - \widetilde{\mathbf P}_{\widetilde{\mathbf A}_r,\boldsymbol \Lambda}$. 
Thus, the optimal solution of problem~\eqref{eqn: GSRxBFProb} is $\mathbf w^{GS}$ given in~\eqref{eqn:w_GS}.

The beamformer $\mathbf w^{GS}$ suggests the following detector
\begin{align} \label{eqn: TGS}
    & T^{GS}(\mathbf y) = \frac{2}{\sigma^2}\frac{\left|\left(\mathbf a_{t} \otimes  (\widetilde{\mathbf P}_{\widetilde{\mathbf A}_r,\boldsymbol \Lambda}^{\perp}\mathbf a_{r})\right)^H\mathbf y\right|^2}{M \mathbf a_{r}^H \widetilde{\mathbf P}_{\widetilde{\mathbf A}_r,\boldsymbol \Lambda}^{\perp} \mathbf a_{r}}.
\end{align}
Because $T^{GS}(\mathbf y)$ uses the Rx-side interference information $\widetilde{\mathbf A}_{r}$ and the Tx-side interference information $\boldsymbol \Lambda$, we call the detector $T^{GS}(\mathbf y)$ as the generalized subspace-based (GS) detector.
From~\eqref{eqn: TGS}, the interference is mitigated using the Rx array, which is the same as the RS detector.
Thus, the GS detector works when the number of interference $Q \leq N$.

The GS detector achieves a balance between interference mitigation gain and object correction gain.
After Rx beamforming, the $q$-th interference residual is
\begin{align} \label{eqn: GSIntProj}
    \left(\mathbf a_{t} \otimes  (\widetilde{\mathbf P}_{\widetilde{\mathbf A}_r,\boldsymbol \Lambda}^{\perp}\mathbf a_{r})\right)^H(\widetilde {\mathbf a}'_{t,q} \otimes \widetilde{\mathbf a}_{r,q})  = \widetilde b_q M \mathbf a_{r}^H \widetilde{\mathbf P}_{\widetilde{\mathbf A}_r,\boldsymbol \Lambda}^{\perp} \widetilde{\mathbf a}_{r,q},
\end{align}
and the object correlation gain is
\begin{align}
    \left(\mathbf a_{t} \otimes  (\widetilde{\mathbf P}_{\widetilde{\mathbf A}_r,\boldsymbol \Lambda}^{\perp}\mathbf a_{r})\right)^H (\mathbf a_{t} \otimes  \mathbf a_{r}) = M \mathbf a_{r}^H \widetilde{\mathbf P}_{\widetilde{\mathbf A}_r,\boldsymbol \Lambda}^{\perp} \mathbf a_{r}.
\end{align}

\subsection{Theoretical Performance Analysis}
\begin{Thm}\label{Thm: TGS}
Based on the assumption $\widetilde b_q \sim \mathcal {CN}(0, h_q^2)$ with known $h_q^2, q=1,2,\ldots, Q$, the probabilities of false alarm and detection for the GS detector under problem~\eqref{eqn: IntHyp} are given as
\begin{align} \label{eqn: pfapdTGS}
    P_{FA}^{GS} = e^{-\frac{1}{2}\gamma}, \quad  P_{D}^{GS} = Q_1 \left(\sqrt{\lambda^{GS}},\sqrt{\gamma}\right),
\end{align}
where $\gamma$ is the detection threshold and 
\begin{align} \label{eqn: lambdaTGS}
    & \lambda^{GS} = \frac{2|b|^2}{\sigma^2} M \mathbf a_{r}^H \widetilde{\mathbf P}_{\widetilde{\mathbf A}_r,\boldsymbol \Lambda}^{\perp} \mathbf a_{r}.
\end{align}
\end{Thm}
\begin{IEEEproof}
See Appendix~B.
\end{IEEEproof}

From the above closed-form expressions of probabilities of false alarm, we have the following Corollary:
\begin{Cor}
The proposed GS detector is a constant false alarm rate (CFAR) detector in the existence of MIMO-FMCW mutual interference. 
\end{Cor}
\begin{IEEEproof}
This CFAR property is ensured by the last condition in problem~\eqref{eqn: GSRxBFProb}, i.e., $(\mathbf P_{\mathbf a_{t}}^{\perp} \otimes  \widetilde{\mathbf a}_{r,q})^H\mathbf w = \mathbf 0_{M}, q=1,2,\ldots,Q$, and the knowledge of $\mathbf R$.
\end{IEEEproof}

\begin{Cor}
The proposed GS detector reduces to the clairvoyant detector of \eqref{eqn: clairvoyant} when the interference Tx steering vectors ${\widetilde {\mathbf a}'}_{t, q}$ are orthogonal to the object Tx steering vector $\mathbf a_{t}$. 
\end{Cor}
\begin{IEEEproof}
The orthorgonality between ${\widetilde {\mathbf a}'}_{t, q}$ and $\mathbf a_{t}$ implies that the EINR matrix $\boldsymbol \Lambda = \mathbf 0$ in \eqref{eqn:lambda} and $\widetilde{\mathbf P}_{\widetilde{\mathbf A}_r,\boldsymbol \Lambda}^{\perp} = \mathbf I$. As a result,  
\begin{align}
        T^{GS}(\mathbf y) = T^{C}(\mathbf y)  
        = \frac{2 |\left(\mathbf a_{t} \otimes  \mathbf a_{r}\right)^H\mathbf y|^2}{\sigma^2 M N }
\end{align}
with $\lambda^{GS} = \lambda^{C} = {2MN|b|^2}/{\sigma^2}$. 
\end{IEEEproof} 

\begin{Cor}
The proposed GS detector reduces to the RS detector of \eqref{eqn: RS} when the projected interference power along the object Tx steering vector approach to infinity. 
\end{Cor}
\begin{IEEEproof}
In this case, the EINR matrix $\boldsymbol \Lambda \to \text{diag}\left[\infty,\infty,\ldots,\infty\right]$. And we have $\widetilde{\mathbf P}_{\widetilde{\mathbf A}_r,\boldsymbol \Lambda} = \widetilde{\mathbf P}_{\widetilde{\mathbf A}_r} =  \widetilde{\mathbf A}_r (\widetilde{\mathbf A}_r^H\widetilde{\mathbf A}_r)^{-1}  \widetilde{\mathbf A}_r^H$ and $\widetilde{\mathbf P}_{\widetilde{\mathbf A}_r,\boldsymbol \Lambda}^{\perp} = \widetilde{\mathbf P}_{\widetilde{\mathbf A}_r}^{\perp}$. As a result, the proposed GS detector of \eqref{eqn: TGS} reduces to the GS detector of \eqref{eqn: RS} with $\lambda^{GS}=\lambda^{RS}= {2M|b|^2 (\mathbf a_{r}^H \mathbf P_{\widetilde{\mathbf A}_{r}}^{\perp}  \mathbf a_{r})}/{\sigma^2}$.
\end{IEEEproof} 

\begin{Cor}
From the probabilities of false alarm and detection of the clairvoyant in \eqref{pfpd_clairvoyant}, RS in \eqref{pfpd_rs} and the proposed GS detectors in Theorem~\ref{Thm: TGS}, the detection performance is in the order of
\begin{align} 
        P_{D}^{RS} \leq P_{D}^{GS} \leq P_{D}^{C}
\end{align}
 for a given probability of false alarm.
\end{Cor}
\begin{IEEEproof}
It is first noted that, for a given probability of false alarm, the detection threshold $\gamma$ holds the same for all three detectors. Then, from Corollary 2 and Corollary 3, we 
\begin{align} \label{eqn: lambdaGSrelation}
       0 < \lambda^{RS} \leq \lambda^{GS} \leq \lambda^{C}, 
\end{align}
when the diagonal elements of EINR matrix $\mathbf \Lambda$ is no smaller than $0$ and finite. Finally, one realizes that the probability of detection or, equivalently, the Marcum Q-function $Q_1(\sqrt{\lambda}, \sqrt{\gamma})$ of order $1$ monotonically increases with respect to $\sqrt{\lambda}$. As a result, the probabilities 
\end{IEEEproof}

\emph{Remark 1}: The GS detector and the LCMV detector are equivalent when the interference statistics are perfectly known. This can be proved by showing $\mathbf w^{GS}$ in~\eqref{eqn:w_GS} is also the optimal solution of the LCMV beamforming optimization problem~\eqref{eqn: LCMVBFProb}. However, the GS detector only needs to estimate EINRs for the interference Tx statistics, while the LCMV detector needs to estimate higher dimensional of unknowns in $\widetilde {\mathbf R}$.

\emph{Remark 2}: When the interference statistics need to be estimated, the GS detector is more robust to the interference statistics estimation error compared to the LCMV detector, because the number of statistics to be estimated for the GS detector is much smaller. We will validate this property using simulation in the next section.

\section{Performance Evaluation}
In this section, simulation results are provided to demonstrate the performance of the proposed GS detector under incoherent MIMO-FMCW mutual interference. We compare the proposed detector of \eqref{eqn: TGS} with 
\begin{itemize}
    \item Clairvoyant detector of \eqref{eqn: clairvoyant},
    \item RS detector of \eqref{eqn: RS} \cite{SAM22},
    \item LCMV detector of \eqref{eqn: TLCMV} \cite{EuRADMIMO},
\end{itemize}
in two simulation scenarios: 
\begin{itemize}
    \item{Synthetic data}:  only the spatial-domain object and interference steering vectors are \emph{directly} synthesized according to the model derived in Section~\ref{sec:model}. More specifically, the object steering vector is generated according to \eqref{ys}, while the interference steering vector is directly generated using \eqref{eqn: decodedIntVirtual}.
    \item{Realistic data}: the received object and interference waveforms go through all necessary steps (LO, LPF, ADC, Rang/Doppler FFT, MIMO waveform separation) at the victim Rx of Fig.~\ref{fig:mimoBlockDiagram} with the help of the MATLAB Phased Array System Toolbox. The realistic data  further accounts for Tx/Rx antenna beampatterns, waveform residuals due to the LPF and imperfect MIMO waveform separation due to the object Doppler modulation, and the noise contributed from spectrum leakage due to the presence of other objects and interferences. 
\end{itemize}

\begin{figure*}[!t]
   \begin{center}
   \begin{tabular}{cc}
   \includegraphics[width=.475\linewidth]{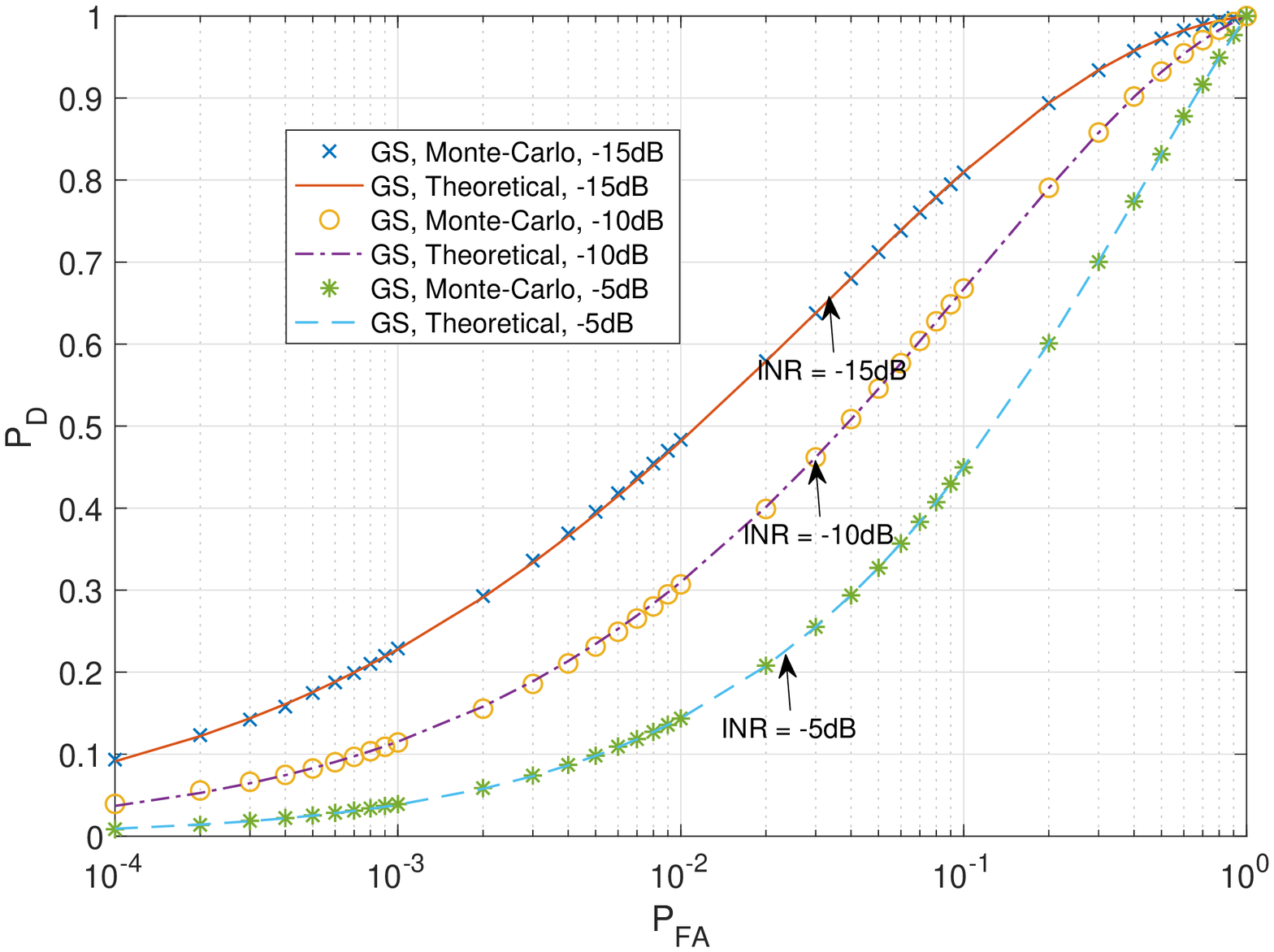} &
   \includegraphics[width=.475\linewidth]{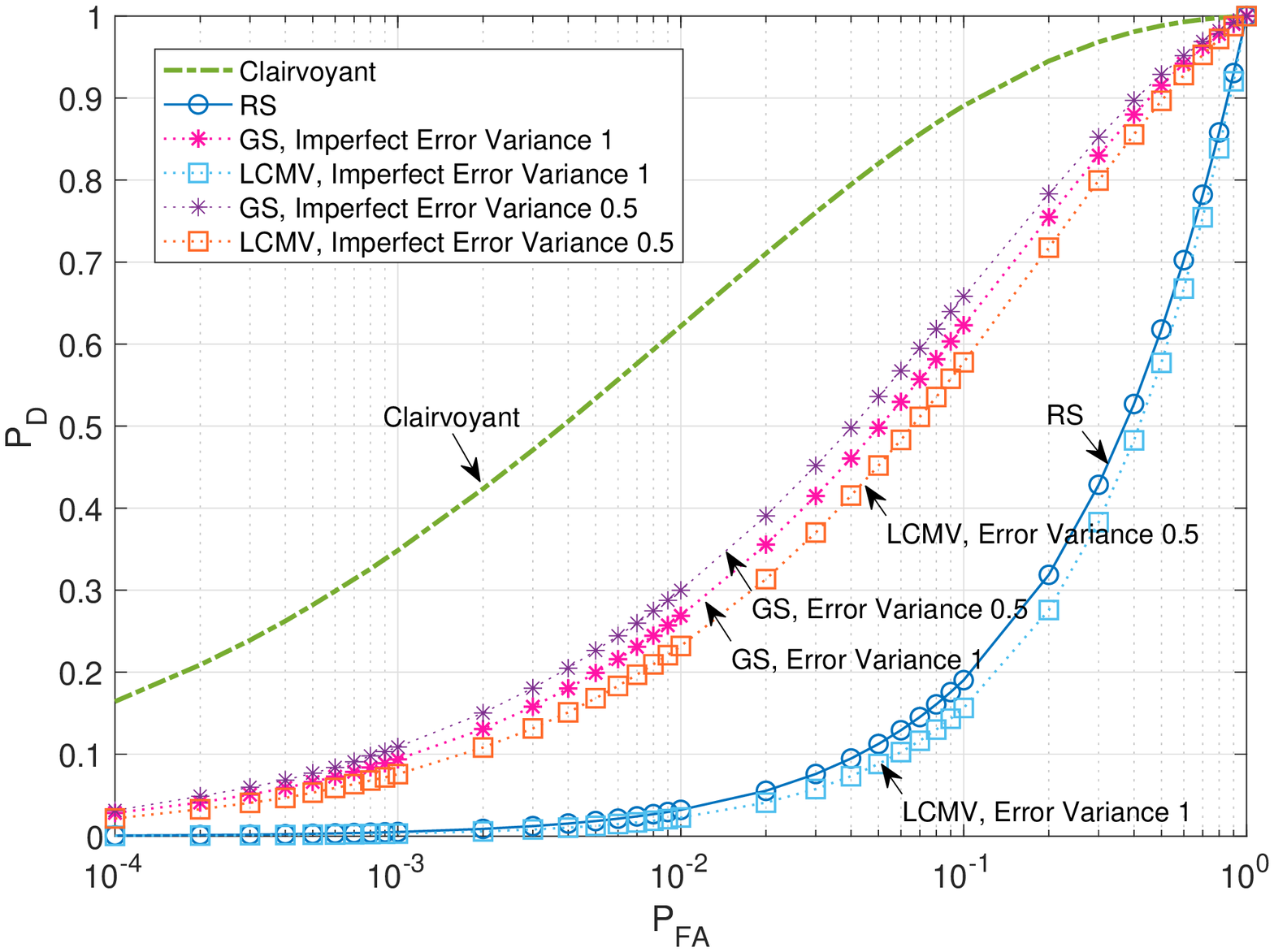}
   \\ (a)&(b)
   \end{tabular}
   \caption{Performance evaluation using synthetic data: Receiver operating characteristic (ROC) curves when $M=4$, $N=4$, and SNR $=-5$ dB in the presence of an object at $30^{\circ}$  and two interferences at $40^{\circ}$ and $10^{\circ}$: (a) Comparison of theoretical (lines) and empirical Monte-Carlo (markers) ROC curves of the proposed GS detector when INR$=\{-15, -10, -5\}$ dB; (b) Empirical Monte-Carlo comparison between the proposed GS detector and baseline methods under two levels of covariance matrix estimation errors. } 
   \label{fig: ROC2}
   \end{center}
\end{figure*}

\subsection{Performance Evaluation using Synthetic Data}

We consider a victim MIMO-FMCW radar with $M=4$ Tx antennas and $N=4$ Rx antennas. The inter-element spacing values at the victim Rx and Tx are $d_r = 0.5\lambda$ and $d_t = N d_r$, respectively. We generate the spatial-domain object steering vector of \eqref{ys} by feeding an object angle at $\phi_t=\phi_r=30^{\circ}$ to the object Tx and Rx steering vectors, respectively. 

At the same time, we consider two mutually independent MIMO-FMCW interferences: one is located at $40^{\circ}$, and the other at $10^{\circ}$ as seen by the victim Rx. We first construct the interference Rx steering vector $\widetilde{\mathbf a}_{r, q}$ according to \eqref{inteferingRxVec} using the two interference angles. For the interference Tx steering vector, since it is incoherent and we have no prior knowledge about interference Tx, we generate it as a random $M \times 1$ vector pointing to an unknown direction in the $M$-dim subspace 
\begin{align}
\widetilde {\mathbf a}'_{t,q} \sim \mathcal {CN}(\mathbf 0, \widetilde{\sigma}_q^2 \widetilde{\mathbf R}_{t,q}).
\end{align}
Note that the direct and random generation of $\widetilde{\mathbf a}'_{t, q}$ ignores the interference Tx configurations and relative geometry between the interference and victim Rx. It provides a simple and computationally efficient way to emulate the interference Tx steering vector in all possible configurations (FMCW/ array configurations and relative interference-victim geometry) and verify our theoretical performance analysis. In our simulation, we set  $\widetilde{\mathbf R}_{t,q} \triangleq [\widetilde{R}_{q,i,j}]_{i,j=0}^{M-1}=[\rho_q^{|i-j|} ] _{i,j=0}^{M-1}$ with $\rho_1=0.6$ and $\rho_2=0.5$ for the two interferences. We define the SNR as $\text{SNR}={|b|^2}/{\sigma^2}$ and set it at $-5$ dB, while the INR is set as $\text{INR}={\widetilde{\sigma}_q^2}/{\sigma^2}$ dB, where $\sigma^2$ is the noise variance. 

\begin{figure}[b]
\begin{center}
\includegraphics[width=0.45\textwidth]{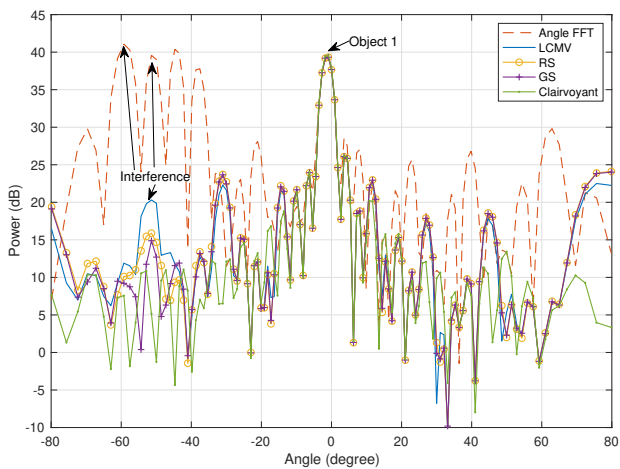}
\end{center}
    \caption{Comparison of angle-domain detection statistics of all considered detectors at a given range-Doppler bin. }
   \label{fig:angleDetection}
\end{figure}

\begin{figure*}[!t]
   \begin{center}
   \begin{tabular}{cc}
   \includegraphics[width=.3\linewidth]{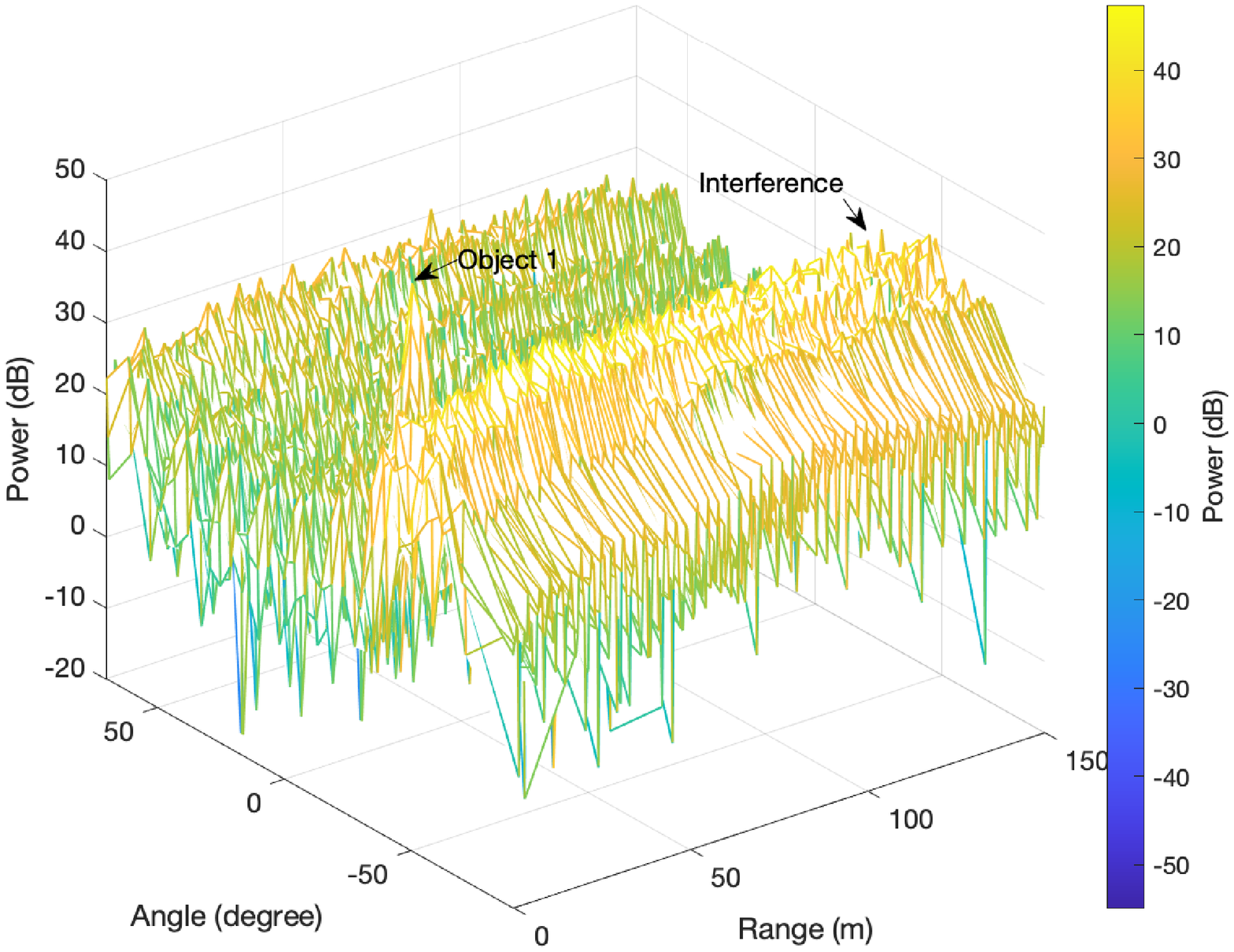} &
   \includegraphics[width=.3\linewidth]{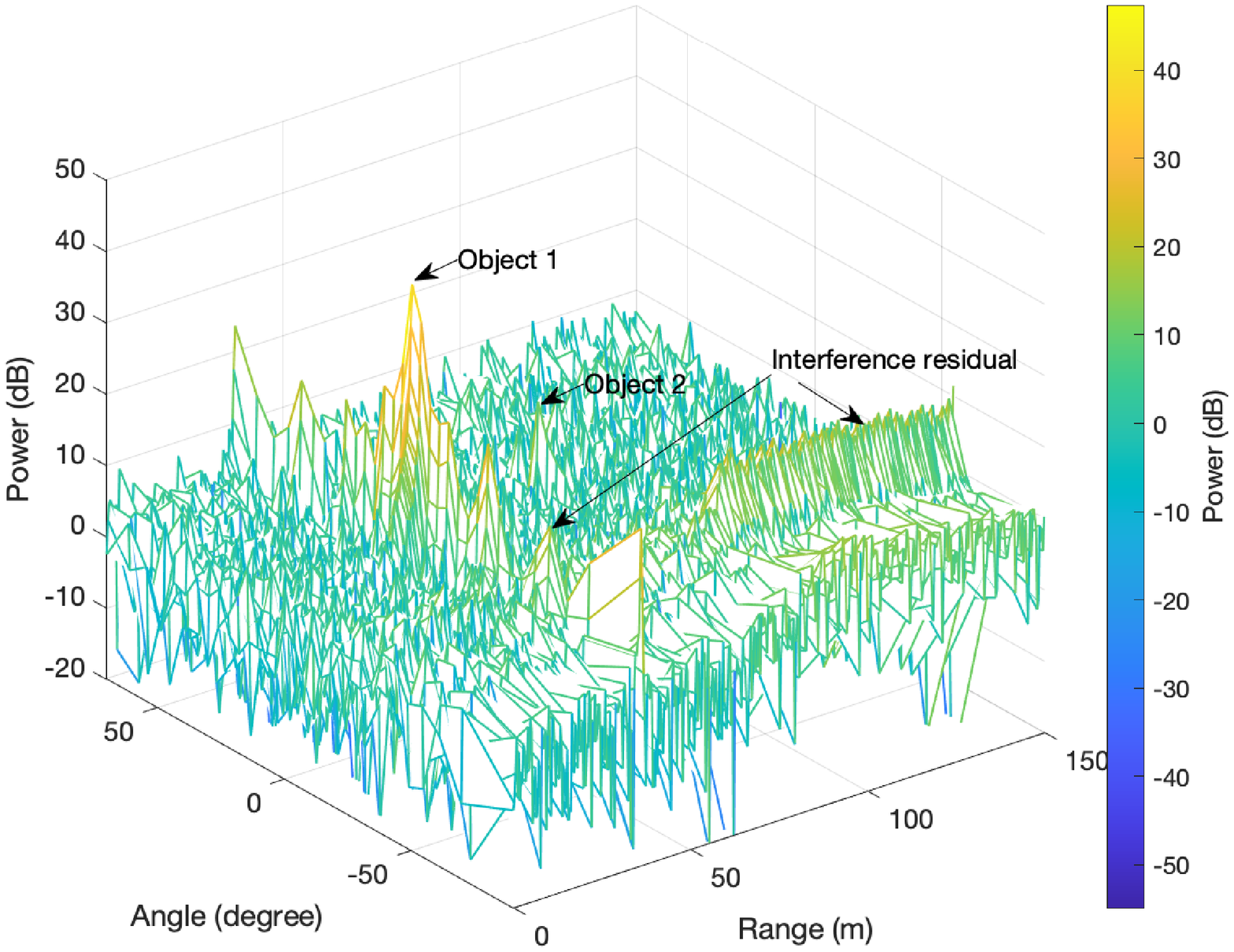} 
    \\ (a) Angular spectrum & (b) LCMV detector
   \end{tabular}
   \begin{tabular}{ccc}
   \includegraphics[width=.3\linewidth]{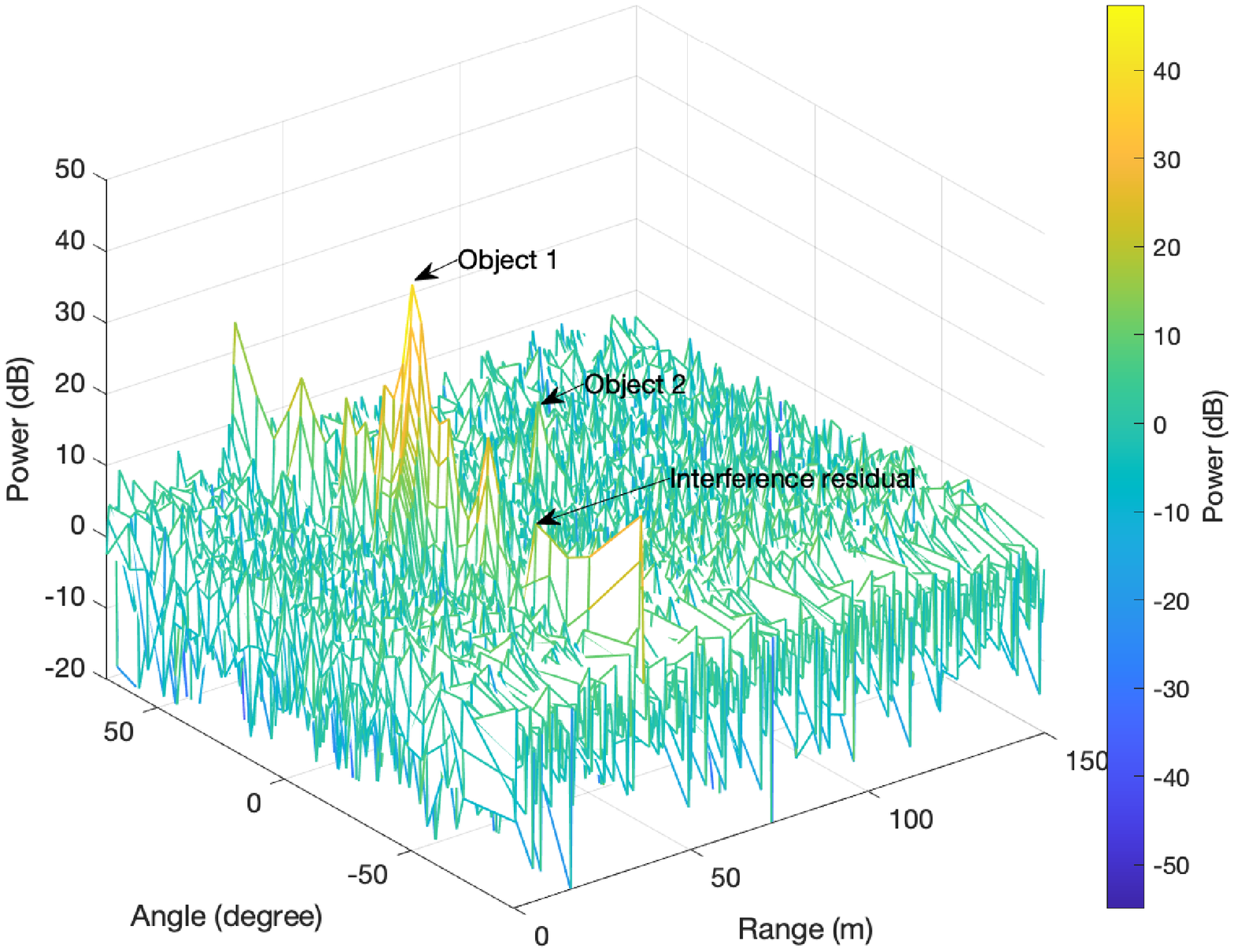} & 
   \includegraphics[width=.3\linewidth]{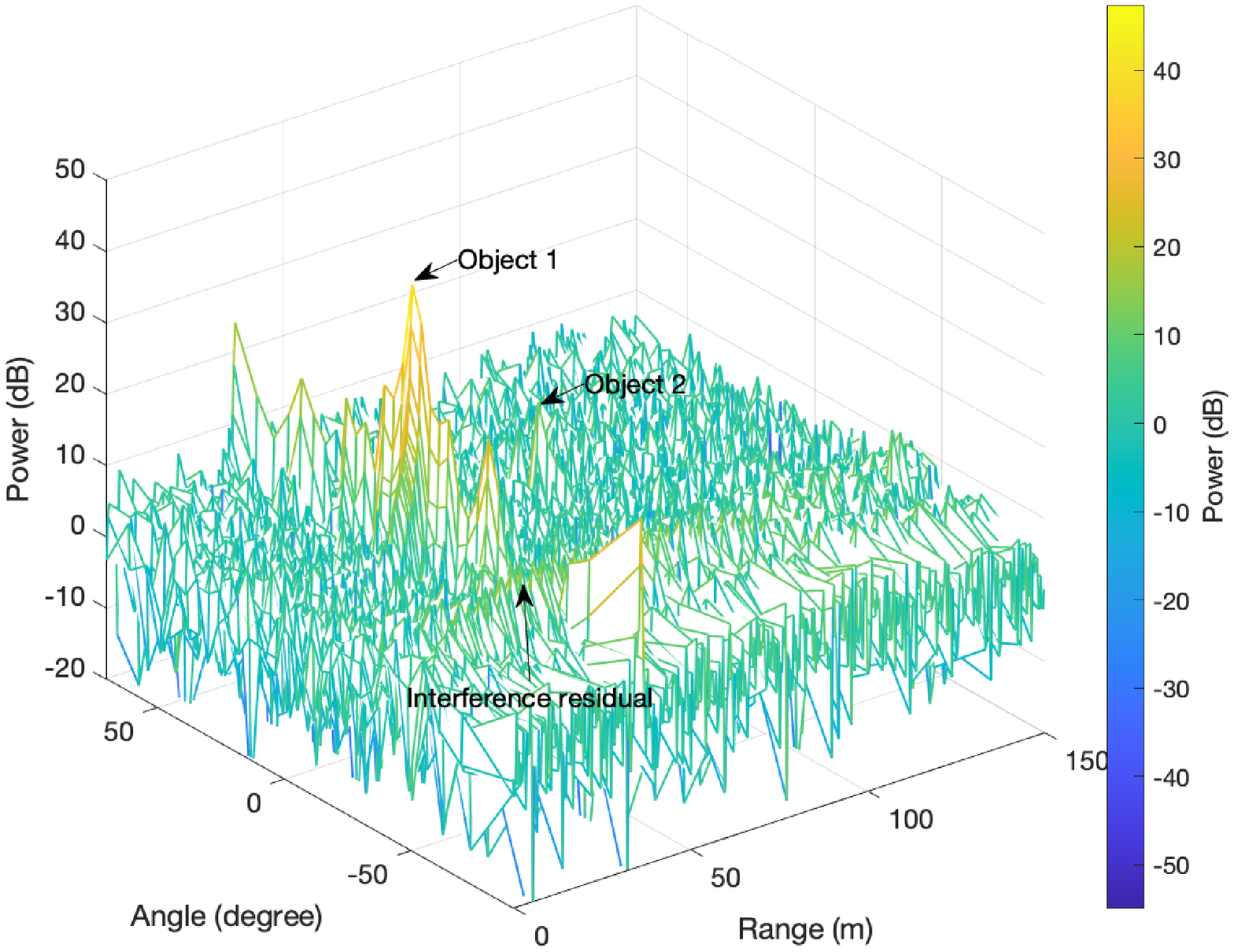} & 
   \includegraphics[width=.3\linewidth]{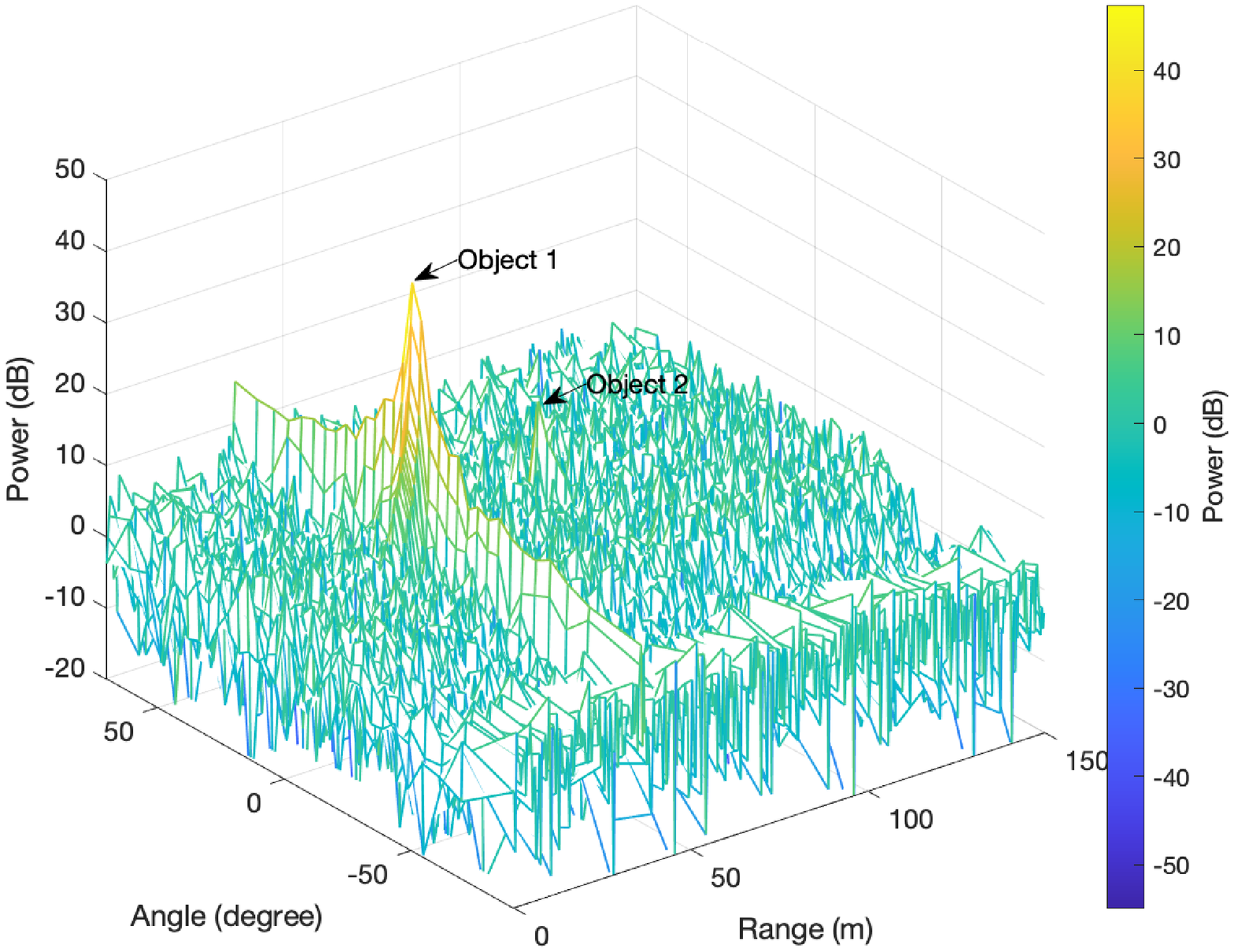} 
    \\ (c) RS detector &(d) Proposed GS detector  & (e) Clairvoyant detector
   \end{tabular}
   \caption{Qualitative detection heatmaps of the proposed detector and baseline detectors in a realistic dataset with $2$ objects and $2$ interferences. All heatmaps are shown over the range-angle domain at the Doppler bin of Object $1$. } 
   \label{fig: toolbox2}
   \end{center}
\end{figure*}

The performance is evaluated in terms of the receiver operating characteristic (ROC) by using Monte Carlo trials. For each Monte-Carlo run, the interference Tx steering vector and noise are randomly generated as specified above, while the interference Rx steering vector and object Tx/Rx steering vectors are fixed according to the specified interference and object angles. 

\begin{table}[t]
 \centering
 \caption{Victim and Interfering MIMO-FMCW Radar Configuration for Realistic Data Generation}\label{tab: phy_setup}
\begin{tabular}{ |c|c|c|c|c|c|c| } 
\hline
Setup  & Explanations\\
\hline
Simulation platform & \makecell{MATLAB Phased \\ Array System Toolbox}\\
\hline
RF wavelength &  $3.9$ mm~\\
\hline
Tx power (Rx noise figure) & $5$ dBm ($4.5$ dB)~\\
\hline
Tx (Rx) antenna gain & $36 (42)$ dB~\\
\hline
Tx (Rx) antenna element type & Backbaffled isotropic ~\\
\hline
Tx (Rx) array structure & Uniform linear array~\\
\hline
MIMO Tx-pulse code & Chu sequence~\\
\hline
Chirp bandwidth & $460$ MHz\\
\hline
IF bandwidth (ADC complex sample rate) & $15$ MHz ($16.7$ MHz)\\
\hline
Number of chirps in a CPI & $256$\\
\hline
Range, velocity, angle FFT sizes & $1024, 256, 32$ \\
\hline
Object RCS model & Non-fluctuating\\
\hline
Object (interference) channel & \makecell{Free-space two-way \\ (one-way) channel} \\
\hline
Victim radar chirp slope & $15$ MHz/us \\
\hline
Victim radar chirp (idle) duration & $30.7$ us ($7$ us) \\
\hline
Victim Tx (Rx) element spacing & $15.6$ mm ($1.95$ mm)~\\
\hline
Victim Tx (Rx) antenna number & $4$ ($8$)\\
\hline
Interfering Tx element spacing & $3.9$ mm  \\
\hline
Interfering Tx antenna number & $8$  \\
\hline
\end{tabular}
\end{table}

We can directly compute the detection statistics of the clairvoyant detector of \eqref{eqn: clairvoyant} with the knowledge of $\widetilde{\mathbf a}_{r, q}$ and $\widetilde{\mathbf a}'_{t, q}$ and the RS detector. On the other hand, the LCMV detector of \eqref{eqn: TLCMV} requires the knowledge of the normalized covariance matrix $\widetilde {\mathbf R}$. To mimic the covariance matrix estimation error due to the lack of homogeneous training data, we use the following perturbed interference Tx covariance matrix \cite{covPer}
\begin{align} \label{interTxPerturb}
     \widetilde {\mathbf R}_{t,q,est} = \widetilde{\sigma}_q^2\widetilde{\mathbf R}_{t,q} \odot (\mathbf 1_M \mathbf 1_M^H + \mathbf E),
\end{align}
where $\mathbf 1_M$ is the all-one vector of dimension $M$, $\mathbf E$ is a $M$-by-$M$ symmetric matrix and each entry in the upper triangular of $\mathbf E$ independently follows zero-mean Gaussian distribution with variance $\sigma^2_{\text{pert}}$, and $\odot$ is the Hadamard product. Consequently, the perturbed normalized covariance matrix of interference-plus-noise used for the LCMV detector is given as
\begin{align} \label{eqn: intTxCov}
    & \widetilde {\mathbf R}_{est} =   \sum_{q=1}^Q \frac{\widetilde{\sigma}_q^2}{\sigma^2} \widetilde{\mathbf R}_{t,q, est} \otimes (\widetilde{\mathbf a}_{r,q}  \widetilde{\mathbf a}_{r,q}^H) + \mathbf I_{MN}.
\end{align}
For a fair comparison, we also use the perturbed interference Tx covariance matrix for the estimation of $h_q^2$ in \eqref{eqn: btuta} as
\begin{align}
    h^2_{q,est} = \frac{\mathbf a_t^H \widetilde{\mathbf R}_{t,q,est} \mathbf a_t}{||\mathbf a_{t}||^4}.
\end{align}

Fig.~\ref{fig: ROC2} (a) verifies the derived theoretical performance (denoted by lines) in \emph{Theorem 1} of Section~\ref{sec: GS} for the proposed GS detector and compares it with empirical ROC curves (denoted by markers) when the INR $=\{-15, -10, -5\}$ dB. A good agreement between the theoretical and empirical ROC curves is observed in Fig.~\ref{fig: ROC2} (a). Second, when the INR decreases or, equivalently, the interference is weaker, the probability of detection increases for a given probability of a false alarm. 

Fig.~\ref{fig: ROC2} (b) further compares the proposed GS detector with the three considered detectors in terms of ROC curves when INR is fixed to $-10$ dB. The clairvoyant detector, although not practical, gives the best detection performance among all detectors. Compared with our previously proposed RS detector, the proposed GS detector shows a significant improvement. For instance, when the probability of false alarm is $0.1$, the probability detection is boosted from $0.2$ of the RS detector to about $0.65$ of the proposed GS detector. When the proposed GS detector is compared with the LCMV detector, we consider two levels of perturbation on the interference Tx covariance matrix in \eqref{interTxPerturb}. Particularly, we consider $\sigma^2_{\text{pert}}=0.5$ and $\sigma^2_{\text{pert}}=1$. When the relatively small perturbation is considered, i.e., $\sigma^2_{\text{pert}}=0.5$, the proposed GS detector is slightly better than the LCMV detector. When the perturbation is increased, the performance of the LCMV detector drops significantly and is even worse than the RS detector. On the other hand, the proposed GS detector maintains its detection performance under the perturbation on the interference Tx covariance matrix.

\subsection{Performance Evaluation using Realistic Data}
In the above synthetic performance evaluation, the object and interference steering vectors were directly generated without considering the presence of Tx/Rx antenna beampatterns, and the presence of waveform residuals due to the LPF and waveform separation. In the following, we consider a system-level performance evaluation by generating the source FMCW-MIMO waveforms at both victim and interference Tx sides, accounting for antenna beampatterns, and including all steps at the victim Rx sides with the help of the MathWorks Phased Array System Toolbox. Particularly, in Table~\ref{tab: phy_setup}, we specify the MIMO-FMCW radar configuration for both victim and interfering radar to synthesize the object and interference steering vectors. 

\begin{table}[t]
 \centering
\caption{Objects and interferer setup.}\label{tab: 2targetsRA}
\begin{tabular}{ |c|c|c|c|c|c|c| } 
\hline
Setup  & Explanations\\
\hline
Objects' RCS & $20$dBsm \\
\hline
Object 1's distance, velocity, angle & $35.5$ m, $-2.9$ m/s, $-1.2^{\circ}$ \\
\hline
Object 2's distance, velocity, angle & $81.0$ m, $4.2$ m/s, $11.2^{\circ}$ \\
\hline
Interferer 1's distance, velocity, angle & $1.8$ m, $1.3$ m/s, $-54.0^{\circ}$ \\
\hline
Interferer 2's distance, velocity, angle & $2.3$ m, $-12.8$ m/s, $-48.1^{\circ}$ \\
\hline
Interferer 1 (2)'s chirp slope & $14.6$ MHz/us ($12.4$ MHz/us)\\
\hline
Interferer 1 (2)'s chirp duration & $31.6$ us ($37.2$ us)\\
\hline
Interferer 1 (2)'s inter-chirp idle duration & $7.5$ us ($7.3$ us)\\
\hline
\makecell{Initial time offset between  \\ victim  radar and interferer 1 (2)} & $20.8$ us ($17.6$ us)\\
\hline
\end{tabular}
\end{table}

We consider a scenario of $2$ objects and $2$ interfering radars in Table~\ref{tab: 2targetsRA}. The detection statistics for all detectors are computed at the range bin of $517$ and the Doppler bin of $128$. Similar to the synthetic data case, the LCMV and proposed GS detectors require the estimation of the noise and interference statistics at the test range-Doppler bin. In this case, we adaptively estimate those statistics from neighboring range-Doppler bins. More specifically, we choose the range-Doppler bins in the range of $\widetilde {\mathcal L} = \{519,515\}$ and $\widetilde {\mathcal K} = \{126,130\}$ with a one-side guard interval of $2$ bins. The implemented adaptive estimation steps are listed in Appendix~\ref{app: adapEst}.

Fig.~\ref{fig:angleDetection} shows \emph{qualitative} detection statistics over the one-dimensional angle at the specified range-Doppler bin. We also include a simple angular-domain FFT by ignoring the presence of mutual interference. It is seen that the interference-ignoring angle FFT yields strong sidelobes around the vicinity of the two interference angles at $-48.1^\circ$ and, respectively, $-54^\circ$. All other detectors show interference mitigation capability at the two interference angles. The LCMV detector shows a relatively stronger sidelobe around these angles potentially due to the interference-plus-noise estimation error. The RS and proposed GS detectors show better interference mitigation at this region of interference angles, while the clairvoyant shows significantly fewer sidelobes over all angles.

Fig.~\ref{fig: toolbox2} show two-dimensional (2D) detection statistics of all detectors by varying both angle and range bins while fixing the Doppler bin at $128$. In Fig.~\ref{fig: toolbox2} (a) of the angle FFT, it is seen that the interference is a wideband signal over the range bins due to the resulting interference at the victim Rx is a chirp-like signal and, hence, it significantly raises the noise level over the detection statistics in the range-angle domain. 
On the other hand, the LCMV detector of Fig.~\ref{fig: toolbox2} (b) shows an improved detection heatmap with smaller sidelobes, lower noise floors, and suppressed interferences around their angles. Stronger interference residuals show up at larger range values, e.g., larger than $100$ m, around the angle of $-50^\circ$. 
Fig.~\ref{fig: toolbox2} (c) and (d) show the 2D detection statistics of the RS and, respectively, the proposed GS detectors. They both show improved interference mitigation than the LCMV detector at larger distances. The proposed GS detector is further better than the RS detector at smaller distances, particularly on the range bin of Object $1$. 
Finally, the clairvoyant detector of Fig.~\ref{fig: toolbox2} (e) provides the best benchmark performance and cancels two interferences completely.

\begin{figure}[t]
\begin{center}
  {\resizebox{8cm}{!}{\includegraphics{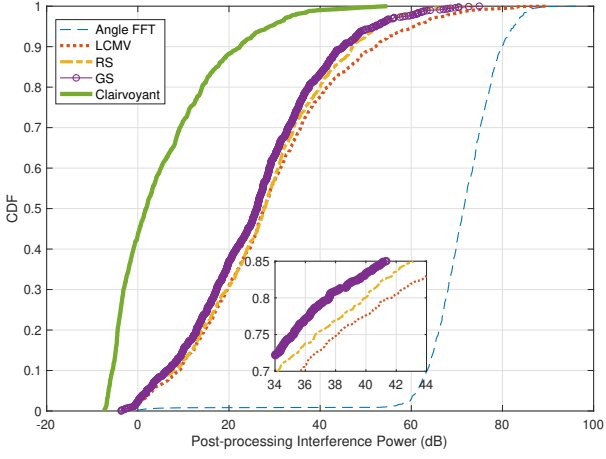}}}
         \caption{CDF of output interference power at the true interference range-angle bin over $1000$ Monte-Carlo runs with a realistic setting.}  \label{fig: intPowCDF}
\end{center}
\end{figure}

We further provide \emph{quantitative} performance evaluation of all considered detection using the realistic data with the Monte-Carlo simulation of $1000$ runs. For each Monte-Carlo run, we randomly select the interference angle in the interval of $[-80^{\circ},80^{\circ}]$ and its range between $1$ m and $3$ m, while keeping other parameters the same as specified in Table~\ref{tab: 2targetsRA}. With randomly selected interference angle and range, we compute the detection statistics at the true interference range-angle bin and refer to it as output interference power (OIP). It is expected that the better the detection performance is, the lower the OIP at the true interference range-angle bin. Fig.~\ref{fig: intPowCDF} shows the cumulative distribution function (CDF) of the OIP of all detectors. It is seen that the proposed GS detector outperforms the RS and LCMV detectors with smaller output interference powers. For instance of the zoom-in window, at the $80$th percentile of the CDF, the output interference powers of the RS and LCMV detectors are $2.5$ dB and $4$ dB higher than that of the proposed GS detector. In other words, the proposed GS detector has better mitigation capability with smaller sidelobes over the interference angles. The clairvoyant detector shows significantly smaller output interference power than the GS detector which implies further improvements are needed to reduce the performance gap. 

\section{Conclusion}
We investigated mutual radar interference mitigation from incoherent MIMO-FMCW automotive radar. By deriving an explicit expression of the incoherent MIMO-FMCW interference that accounts for FMCW incoherence as well as MIMO array differences, we formulated the mutual interference mitigation as a spatial-domain object detection under Kronecker structured interference. Compared with existing spatial-domain detectors, e.g., the LCMV and RS detectors, we proposed a GS detector that exploits the structure of both transmit and receiver steering vectors of the incoherent interference and 
proved that it is a CFAR detector. Both analytical and empirical performance evaluations using directly generated synthetic data and realistic data with the help of the phased array toolbox confirmed the performance improvements in terms of detection performance and output interference power.

\begin{appendices}

\section{Derivation of Object Signal Model}
\label{app: object}
In the following, we show the detaied derivation of the object signal model following the steps in the upper right of Fig.~\ref{fig:mimoBlockDiagram}.  

\emph{Local Oscillator (LO)}: At the $n$-th Rx antenna of the victim radar, the backscattered object signal  $\alpha \sum_{m=0}^{M-1} s_m(t - \tau_{m,n} (t))$ is mixed with the conjugate of the LO signal $\sum_{k=0}^{K-1} s^*(t-k T_{\text{PRI}}) e^{-j2\pi f_c (t -k T_{\text{PRI}})}$, leading to the dechirped baseband analog signal 
\begin{align}
   a^s_n(t)= &\alpha_{\tau} \sum_{m=0}^{M-1} e^{-j2\pi f_c \frac{2vt}{c}} e^{- j2\pi (f_{\phi_t} m + f_{\phi_r} n)} \nonumber\\
   & \times  \sum_{k=0}^{K-1} c_{k,m}  e^{-j2\pi \beta (t-k T_{\text{PRI}})\tau} D_{\tau,T} (t- k T_{\text{PRI}}), 
\end{align}
where $\alpha_{\tau} \triangleq \alpha e^{- j2\pi f_c \tau} e^{j\pi \beta \tau^2}$ with $\alpha$ denoting the complex object amplitude,   
$f_{\phi_t} = d_t {\sin(\phi_{t})}/{\lambda}$ and $f_{\phi_r} = d_r {\sin(\phi_{r})}/{\lambda}$ are the Tx and Rx normalized spatial frequencies at wavelength $\lambda = {c}/{f_c}$, and $\tau = {2R}/{c}$ is the round-trip propagation delay at the $0$-th Rx antenna (reference antenna). 

\emph{Analog-to-Digital Converter (ADC) and Low-Pass Filter (LPF)}: 
Suppose the object beat frequency $\beta \tau$ is smaller than the cutoff frequency $f_L$ of the anti-aliasing LPF.
By passing $a^s_n(t)$ into the LPF and sampling it at $t = kT_{\text{PRI}} + l \Delta T$ with $\Delta T$ denoting the fast-time interval, we have the sampled object signal on fast-time sample $l$ and pulse $k$, i.e., 
\begin{align}
       a^s_{n}(l,k) = & \alpha_{\tau} e^{-j2\pi f_r l} \mathbf{1}[l \in \mathcal L^s]  \nonumber\\
      \times & \sum_{m=0}^{M-1} c_{k,m} e^{- j2\pi (f_d k+ f_{\phi_t} m + f_{\phi_r} n)},
\end{align}
where $\mathcal L^s \triangleq \{\lceil {\tau}/{\Delta T} \rceil,\ldots, \lfloor {T}/{\Delta T}\rfloor\}$ is the set of integer sample indices, $f_r \triangleq (\beta \tau + {2v}/{\lambda})\Delta T$ is the normalized range frequency, and $f_d \triangleq 2f_c T_{\text{PRI}} v /c$ is the normalized Doppler frequency. 

\emph{Fast-Time/Range FFT}: Applying the $L$-length fast-time fast Fourier transform (FFT) or range FFT to $a^s_{n}(l,k)$, we can obtain the range-domain spectrum as
\begin{align} \label{eqn: x_n_s}
 x^s_{n}(l',k) =  \alpha_{l'} \sum_{m=0}^{M-1} c_{k,m}  e^{-j2\pi f_d k} e^{-j2\pi (f_{\phi_t} m + f_{\phi_r} n)},  
\end{align}
where $\alpha_{l'} \triangleq \sum_{l=0}^{L-1} \alpha_{\tau} \mathbf{1}[l \in \mathcal L^s] e^{-j2\pi (f_r+{l'}/{L})l}$
is the complex amplitude of the object on range bin $l'$. 

\emph{Slow-Time/Doppler FFT and Waveform Separation}:
From~\eqref{eqn: x_n_s}, each Rx antenna combines the $M$ coded transmitting waveforms via the weighted summation. 
To separate $x^s_{n}(l',k)$ into object signals from $M$ Tx signals, a slow-time MIMO decoding is applied.
To obtain the signal from $m$-th Tx antenna, the complex conjugate of the code sequence $c_{k,m}^*, k = 0,1,\ldots, K-1$ are multiplied on the range-domain response over $K$ slow-time pulses.
For a MIMO code sequence with orthogonal property $\sum_{k=0}^{K-1} c_{k,m} c_{k,m}^* = K, \sum_{k=0}^{K-1} c_{k,m} c_{k,m'}^* = 0, \forall \ m' \neq m$, summing the decoded signal over $K$ pulses $\sum_{k=0}^{K-1} x^s_{n}(l',k) c_{k,m}^*$ can well reconstruct the object signal with zero Doppler from $m$-th Tx antenna.
For a general case where the slow-time phase is shifted by the non-zero object Doppler, the Doppler needs to be compensated.
To reconstruct the object signal from the $m$-th Tx antenna, we can compensate the Doppler using a slow-time FFT (Doppler FFT) on the slow-time decoded signal $x^s_{n}(l',k) c_{k,m}^*, k=0,1,\ldots,K-1$:
\begin{align} \label{eqn: RDySignal}
 y^s_{m,n}(l',k') = & \sum_{k=0}^{K-1} x^s_{n}(l',k) c_{k,m}^* e^{-j2\pi \frac{k'}{K}k} \\
    = &  b(l',k') e^{-j2\pi (f_{\phi_t} m + f_{\phi_r} n)} + 
   y^r_{m,n}(l',k'), \notag 
\end{align}
where $b(l',k') \triangleq \alpha_{l'} \sum_{k=0}^{K-1}  e^{-j2\pi (f_d + \frac{k'}{K}) k}$ is the amplitude of the object signal from the $m$-th Tx antenna, and 
\begin{align}
y^r_{m,n}(l',k') =   \alpha_{l'} \underset{m' \neq m}\sum & \left(\sum_{k=0}^{K-1} c_{k,m'}  c_{k,m}^* e^{-j2\pi (f_d + \frac{k'}{K}) k} \right) \notag \\
& \times e^{-j2\pi (f_{\phi_t} m' + f_{\phi_r} n)},
\end{align}
is the waveform separation residual from other Tx antennas.  
At the Doppler bin $k'$ closest to the object Doppler frequency $f_d$, i.e., $f_d + {k'}/{K} \approx 0$, the amplitude $b(l',k') \approx K \alpha_{l'}$ approaches to a coherent gain of $K$ due to the Doppler FFT in \eqref{eqn: RDySignal}.
Using the near-orthogonality of MIMO codes~\cite{SunPetropulu20}
\begin{align} \label{eqn: orthogonality}
    \max_f \left|\sum_{k=0}^{K-1} c_{k,m} c_{k,m'}^* e^{-j2\pi f k}\right| \ll K, \forall \ m' \neq m,
\end{align}
the waveform separation residual in \eqref{eqn: RDySignal} can be ignored. It is worthy noting that object detection under imperfect waveform separation for MIMO radar has been considered in \cite{WangBoufounos20} and \cite{WangLi20}.

\section{Derivation of Interference Signal Model}
\label{app: interference}
\begin{figure}[t]
\begin{center}
  {\resizebox{7.5cm}{!}{\includegraphics{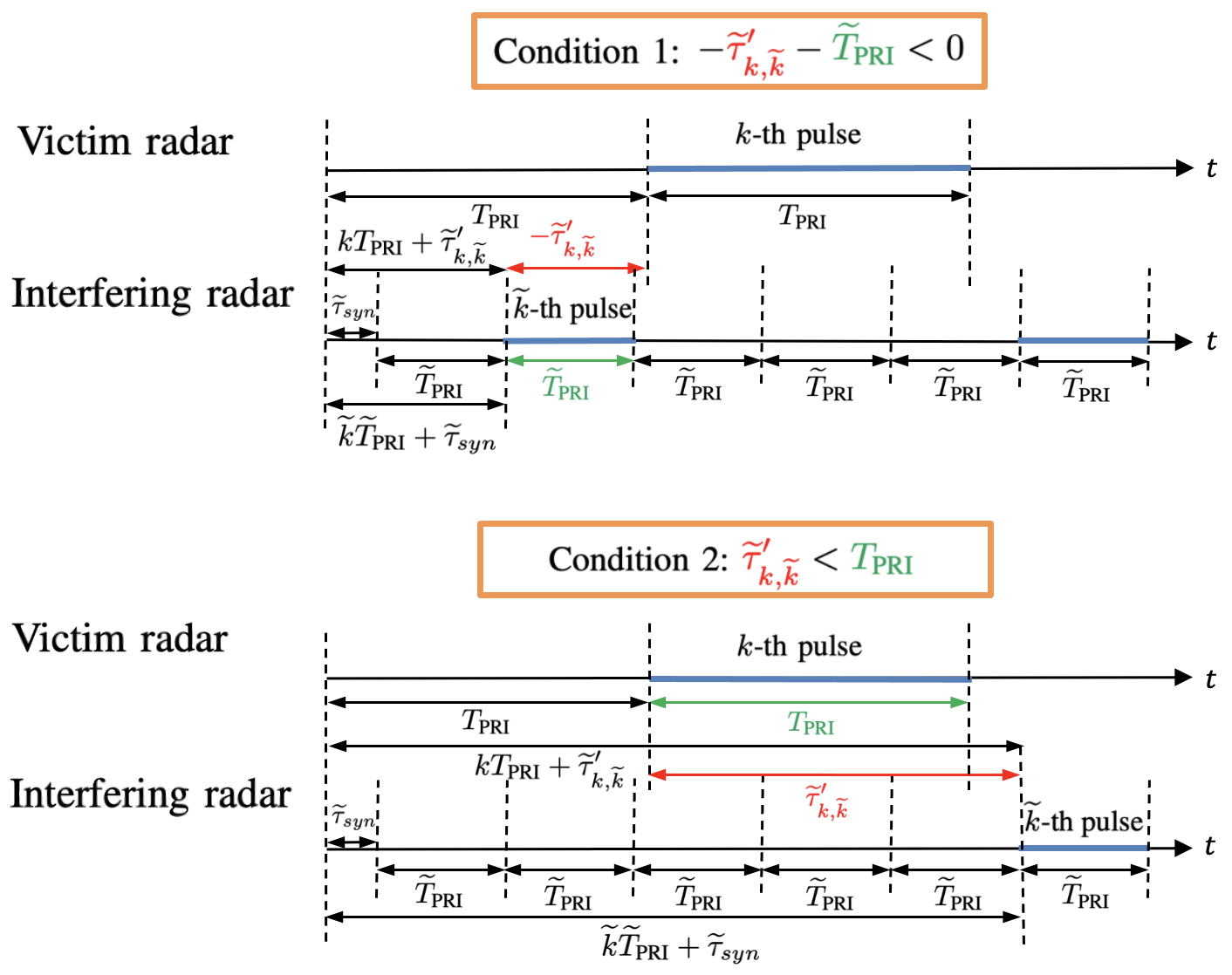}}}
         \caption{\small{Two necessary conditions for the $\widetilde k$-th pulse of the interfering radar to be dechirped by the $k$-th pulse of the victim radar with a counterexample for each condition shown in the figure.
          }}\label{fig:IntSlowTimeCond}
\end{center}
\vspace{-0.1in}
\end{figure}

Two necessary conditions for the $\widetilde k$-th pulse of the interfering radar be dechirped by the $k$-th pulse of victim radar are 
$-\widetilde \tau'_{k,\widetilde k} - \widetilde T_{\text{PRI}} < 0$ and $\widetilde \tau'_{k,\widetilde k} < T_{\text{PRI}}$, which in combination lead to $-\widetilde T_{\text{PRI}} < \widetilde \tau'_{k,\widetilde k} < T_{\text{PRI}}$.
If any of these two conditions does not satisfy, as shown in Fig.~\ref{fig:IntSlowTimeCond}, the $\widetilde k$-th pulse of the interfering radar cannot be dechirped by the $k$-th pulse of victim radar.

Define
\begin{align} \label{eqn: IntKset}
    \mathcal K_{\widetilde k}  \triangleq \Big\{& k:  \widetilde k \widetilde T_{\text{PRI}} + \widetilde \tau_{syn} = k T_{\text{PRI}} + \widetilde \tau'_{k,\widetilde k}, -\widetilde T_{\text{PRI}} < \widetilde \tau'_{k,\widetilde k} < T_{\text{PRI}}, \nonumber \\
   & k = 0,1,\ldots, K-1 \Big\}, \ \widetilde k = 0,1,\ldots, \widetilde K-1,
\end{align}
as a set that groups all pulses of the victim radar that intercept with the $\tilde k$ pulse by checking whether any time instant of the victim pulse falls within the $\tilde k$ interfering pulse. 
Denote the slow-time code of the interfering radar observed at $k$-th victim radar pulse as
\begin{align}\label{eqn: rxIntMIMOcode}
    \widetilde c_{k,\widetilde m}^{\widetilde k} \triangleq
    \begin{cases}
    \widetilde c_{\widetilde k,\widetilde m}, & k \in \mathcal K_{\widetilde k} \\
    0, & \text{otherwise}.
\end{cases}
\end{align}
Then, we rewrite $s^i_n(t)$ as
\begin{align}
    & s^i_n(t) = \widetilde \alpha e^{- j2\pi f_c \widetilde \tau} \sum_{\widetilde m=0}^{\widetilde M-1} \sum_{\widetilde k=0}^{\widetilde K-1} \sum_{ k \in \mathcal K_{\widetilde k}} \widetilde c_{k,\widetilde m}^{\widetilde k} \widetilde s(t-k T_{\text{PRI}} - \widetilde \tau'_{k,\widetilde k} - \widetilde \tau)   \nonumber\\
   & \times e^{j2\pi f_c (t-k T_{\text{PRI}} - \widetilde \tau'_{k,\widetilde k})} e^{- j2\pi (\widetilde f_{\phi_t} \widetilde m + \widetilde f_{\phi_r} n)}  e^{-j2\pi f_c \frac{\widetilde vt}{c}}.
\end{align}
The victim radar mixes $s^i_n(t)$ with its LO signal, and obtains the analog beat signal from the $n$-th Rx antenna
\begin{align}\label{eqn: beforeLPF}
    & a^i_n(t) = s^i_n(t) \sum_{k=0}^{K-1} s^*(t-k T_{\text{PRI}}) e^{-j2\pi f_c (t -k T_{\text{PRI}})}\nonumber\\
   =& \widetilde \alpha e^{- j2\pi f_c \widetilde \tau} \sum_{\widetilde m=0}^{\widetilde M-1} \sum_{\widetilde k=0}^{\widetilde K-1} \sum_{k \in \mathcal K_{\widetilde k}} \widetilde c_{k,\widetilde m}^{\widetilde k} e^{j\pi (\widetilde \beta - \beta) (t - k T_{\text{PRI}})^2}   \nonumber\\
   & \times e^{j\pi \widetilde \beta (\widetilde \tau'_{k,\widetilde k} + \widetilde \tau)^2} 
   e^{-j2\pi \widetilde \beta (t-k T_{\text{PRI}}) (\widetilde \tau'_{k,\widetilde k} + \widetilde \tau)}\nonumber\\
   & \times e^{-j2\pi f_c  \widetilde \tau'_{k,\widetilde k}} e^{- j2\pi (\widetilde f_{\phi_t} \widetilde m + \widetilde f_{\phi_r} n)}  e^{-j2\pi f_c \frac{\widetilde vt}{c}} \nonumber\\
   & \times D_{\widetilde \tau'_{k,\widetilde k} + \widetilde \tau,\min\left\{T, \widetilde \tau'_{k,\widetilde k} + \widetilde \tau + \widetilde T\right\}} (t-k T_{\text{PRI}}).
\end{align}

\begin{figure}[t]
\begin{center}
  {\resizebox{7.5cm}{!}{\includegraphics{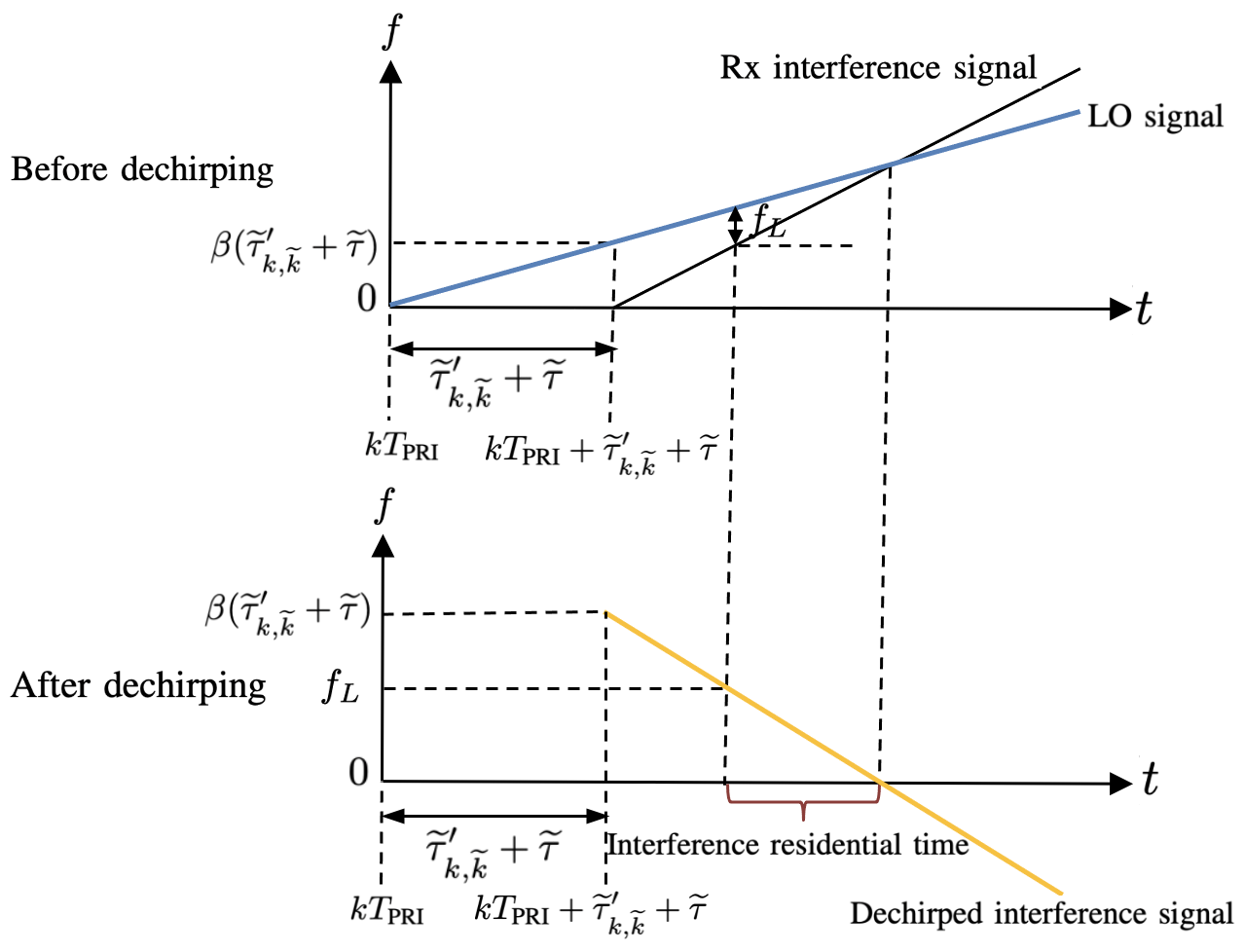}}}
         \caption{\small{Interference at victim radar's pulse $k$.
          }}\label{fig:LPF}
\end{center}
\vspace{-0.1in}
\end{figure} 

From \eqref{eqn: beforeLPF}, we can see that the instantaneous frequency of interference at pulse $k$ is $\widetilde \beta (\widetilde \tau'_{k,\widetilde k} + \widetilde \tau) - (\widetilde \beta - \beta) (t - k T_{\text{PRI}})$.
Then, passing $a^i_n(t)$ into the LPF of bandwidth $f_L$, the interference residential time on pulse $k$ with interference is 
\begin{align} \label{eqn: LPFcondition}
    0<  \widetilde \beta (\widetilde \tau'_{k,\widetilde k} + \widetilde \tau) - (\widetilde \beta - \beta) (t - k T_{\text{PRI}}) < f_L.
\end{align}
Fig.~\ref{fig:LPF} provides an illustrative example showing the interference residential time.
The low-pass filtered IF interference signal is
\begin{align}
    a^{i,low}_n(t) & =  \widetilde \alpha e^{- j2\pi f_c \widetilde \tau} \sum_{\widetilde m=0}^{\widetilde M-1} \sum_{\widetilde k=0}^{\widetilde K-1} \sum_{k \in \mathcal K_{\widetilde k}} \widetilde c_{k,\widetilde m}^{\widetilde k} e^{j\pi (\widetilde \beta - \beta) (t - k T_{\text{PRI}})^2}   \nonumber\\
   & \times e^{j\pi \widetilde \beta (\widetilde \tau'_{k,\widetilde k} + \widetilde \tau)^2} 
   e^{-j2\pi \widetilde \beta (t-k T_{\text{PRI}}) (\widetilde \tau'_{k,\widetilde k} + \widetilde \tau)}\nonumber\\
   & \times e^{-j2\pi f_c  \widetilde \tau'_{k,\widetilde k}} e^{- j2\pi (\widetilde f_{\phi_t} \widetilde m + \widetilde f_{\phi_r} n)}  e^{-j2\pi f_c \frac{\widetilde vt}{c}} \nonumber\\
   & \times \mathbf{1} \left[0<  \widetilde \beta (\widetilde \tau'_{k,\widetilde k} + \widetilde \tau) - (\widetilde \beta - \beta) (t - k T_{\text{PRI}}) < f_L\right]\nonumber\\
   & \times D_{\widetilde \tau'_{k,\widetilde k} + \widetilde \tau,\min\left\{T, \widetilde \tau'_{k,\widetilde k} + \widetilde \tau + \widetilde T\right\}} (t-k T_{\text{PRI}}).
\end{align}

\section{Proof of Theorem~\ref{Thm: TGS}}
\label{app: proof}
The following derivation is based on the form $T^{GS}(\mathbf y) = \frac{2}{\sigma^2}\frac{\left|\left(\mathbf R^{-1}  (\mathbf a_{t} \otimes  \mathbf a_{r})\right)^H\mathbf y\right|^2}{\big|\big| \mathbf R^{-\frac{1}{2}} (\mathbf a_{t} \otimes  \mathbf a_{r})\big|\big|^2}$ suggested by $\mathbf w^{GS}$ in~\eqref{eqn:w_GS}.

Under $H_0$, we have $(\mathbf a_{t} \otimes  \mathbf a_{r})^H \mathbf R^{-1}\mathbf y = (\mathbf a_{t} \otimes  \mathbf a_{r})^H \mathbf R^{-1}\widetilde {\mathbf z}$, using the last condition in \eqref{eqn: GSRxBFProb}. 
As $\widetilde {\mathbf z} \sim \mathcal {CN}(\mathbf 0, \sigma^2 \mathbf R)$ by \eqref{eqn: z_tu_ta}, we have $(\mathbf a_{t} \otimes  \mathbf a_{r})^H \mathbf R^{-1}\mathbf y \sim \mathcal {CN}\left(\mathbf 0, \sigma^2 \Big|\Big| \mathbf R^{-\frac{1}{2}} (\mathbf a_{t} \otimes  \mathbf a_{r})\Big|\Big|^2\right)$. Thus, $T^{GS}(\mathbf y)$ under $H_0$ follows chi-squared distribution with $2$ degrees of freedom (DoF), i.e., 
\begin{align}
    T^{GS}(\mathbf y) \sim \chi_{2}^2, \ \text{under} \ H_0.
\end{align}

Under $H_1$, we have $(\mathbf a_{t} \otimes  \mathbf a_{r})^H \mathbf R^{-1}\mathbf y  \sim \mathcal {CN}\left(b \Big|\Big| \mathbf R^{-\frac{1}{2}} (\mathbf a_{t} \otimes  \mathbf a_{r})\Big|\Big|^2, \sigma^2 \Big|\Big| \mathbf R^{-\frac{1}{2}} (\mathbf a_{t} \otimes  \mathbf a_{r})\Big|\Big|^2\right)$. Thus, $T^{GS}(\mathbf y)$ under $H_1$ follows noncentral chi-squared distribution with $2$ DoF and noncentrality parameter $\lambda^{GS}$, i.e., 
\begin{align}
    T^{GS}(\mathbf y) \sim {\chi'}_{2}^{2}(\lambda^{GS}), \ \text{under} \ H_1,
\end{align}
where $\lambda^{GS} = \frac{2|b|^2}{\sigma^2} \Big|\Big|\mathbf R^{-\frac{1}{2}} (\mathbf a_{t} \otimes  \mathbf a_{r})\Big|\Big|^2 = \frac{2|b|^2}{\sigma^2} M \mathbf a_{r}^H \widetilde{\mathbf P}_{\widetilde{\mathbf A}_r,\boldsymbol \Lambda}^{\perp} \mathbf a_{r}$.
By $P_{FA}^{GS} = \Pr\left[T^{GS}(\mathbf y) \geq \gamma| H_0\right]$ and $P_{D}^{GS} = \Pr\left[T^{GS}(\mathbf y) \geq \gamma| H_1\right]$, we have \eqref{eqn: pfapdTGS}.

\section{Adaptive Estimation of Interference and Noise Statistics in The Case of Realistic Data}
\label{app: adapEst}
Similar to the synthetic data case, the LCMV and GS detectors need the interference-plus-noise covariance matrix and, particularly, the interference Tx covariance matrix to compute the detection statistics. Unlike the synthetic data case, we do not have the access to the true interference covariance and, hence, its perturbation. As a result, we need to adaptive estimate the interference-plus-noise covariance matrix or, equivalently,  $h_q^2$, $\widetilde{\sigma}_q^2\widetilde{\mathbf R}_{t,q}$ and $\sigma^2$, from neighboring range-Doppler bins. 

At an object-free range-Doppler bin $(l',k')$ of the victim radar, the received spatial-domain signal is $\mathbf y (l',k') = \sum_{q=1}^Q \widetilde {\mathbf a}'_{t,q} \otimes \widetilde{\mathbf a}_{r,q} + \mathbf z$. The noise power $\sigma^2$ at the same range-Doppler bin can be estimated as
\begin{align}
    \widehat \sigma^2 (l',k') = \frac{2}{M(N-Q)} || (\mathbf I_M \otimes \mathbf P_{\widetilde{\mathbf A}_{r}}^{\perp}) \mathbf y (l',k')||^2,
\end{align}
the $q$-th interference Tx steering vector $\widetilde {\mathbf a}'_{t,q}$ can be estimated as 
\begin{align}
    \widehat {\mathbf a}'_{t,q} (l',k') = \left(\mathbf I_M \otimes (\widetilde{\mathbf A}_{r} \mathbf b_q)^H\right) \mathbf y (l',k'),
\end{align}
where $\mathbf b_q$ is the $q$-th column of $(\widetilde{\mathbf A}_{r}^H \widetilde{\mathbf A}_{r})^{-1}$, and $\widetilde b_q$ can be estimated as
\begin{align}  
    \widehat b_q (l',k') = \frac{\mathbf a_{t}^H\widehat {\mathbf a}'_{t,q} (l',k')}{||\mathbf a_{t}||^2}.
\end{align}

As a result, by collecting a set of range-Doppler bins, e.g.,  $\widetilde {\mathcal L}$ and $\widetilde {\mathcal K}$, we can average out the noise power estimate as 
\begin{align}
    \widehat \sigma^2 = \frac{1}{|\widetilde {\mathcal L}| |\widetilde {\mathcal K}|} \sum_{l' \in \widetilde {\mathcal L}, k' \in \widetilde {\mathcal K}} \widehat \sigma^2 (l',k'). 
\end{align}
In a similar fashion,, we have
\begin{align}
   \widehat h_q^2 = \frac{1}{|\widetilde {\mathcal L}| |\widetilde {\mathcal K}|} \sum_{l' \in \widetilde {\mathcal L}, k' \in \widetilde {\mathcal K}} |\widehat b_q (l',k')|^2,
\end{align}
and
\begin{align}
    \widehat{\mathbf R}_{t,q} = \frac{1}{|\widetilde {\mathcal L}| |\widetilde {\mathcal K}|} \sum_{l' \in \widetilde {\mathcal L}, k' \in \widetilde {\mathcal K}} \widehat {\mathbf a}'_{t,q} (l',k') \widehat {\mathbf a}'^H_{t,q} (l',k').
\end{align}

\end{appendices}

\bibliography{reference}

\end{document}